\pdfoutput=1
\documentclass[a4paper,twocolumn]{article}
\usepackage[left=2cm,right=1cm,top=2cm,bottom=2cm]{geometry}

\usepackage{amsmath}
\usepackage{amssymb}
\usepackage{enumitem}
\usepackage{tabularx}
\usepackage{ltxtable}
\usepackage{longtable}
\usepackage{booktabs}
\usepackage{bbold}
\usepackage{braket}
\usepackage{mathtools}
\usepackage{cuted}
\usepackage{times}
\usepackage{authblk}

\vfuzz=\maxdimen
\hfuzz=\maxdimen
\usepackage{tikz,graphicx,etoolbox}
\usetikzlibrary{decorations.markings, calc, positioning, decorations.pathmorphing, fixedpointarithmetic,shapes,patterns}
\AtBeginEnvironment{tikzpicture}{\catcode`$=3 } 
\tikzset{alignmid/.style={baseline={([yshift=-.5ex]current bounding box.center)}}} 
\tikzset{every picture/.append style=alignmid}

\usepackage{hyperref}
\definecolor{darkblue}{RGB}{46,48,146}
\hypersetup{colorlinks=true,linkcolor=darkblue,pdfborderstyle={0 0 0}}

\usepackage{amsthm}
\theoremstyle{definition}

\allowdisplaybreaks[1]

\usepackage{ifthen}
\usepackage{xparse,pgffor}
\ExplSyntaxOn

\NewDocumentCommand{\Foreach}{ m O{,} m}
 {
  \__foreach_main:nn { #1 } { #2 } { #3 }
 }

\seq_new:N \l__foreach_arg_seq
\clist_new:N \l__foreach_braced_clist

\cs_new_protected:Npn \__foreach_main:nn #1 #2 #3
 {
  \seq_set_split:Nnn \l__foreach_arg_seq { #2 } { #3 }
  \clist_clear:N \l__foreach_braced_clist
  \seq_map_inline:Nn \l__foreach_arg_seq
   {
    \clist_put_right:Nn \l__foreach_braced_clist { { ##1 } }
   }
  \foreach #1 in~\l__foreach_braced_clist
 }
 \ExplSyntaxOff

\newcommand\coords[1]{\Foreach{\x/\y}[|]{#1}{\coordinate (\x) at (\y);}}
\newcommand\shapes[2]{\foreach \x in {#2}{#1{(\x)}}}

\tikzset{dec/.style= {postaction=decorate,decoration={name=markings,mark=at position 0.5 with{#1{(0,0)}}}}}

\tikzset{enddots/.style= {postaction=decorate,decoration={name=markings,mark=at position 1 with{
    \ifdimless{\pgflinewidth}{1pt}{\newcommand\dotlw{1pt}}{\newcommand\dotlw{\pgflinewidth}}
    \ifdimless{\pgflinewidth}{1.5pt}{\newcommand\dotdist{1.5pt}}{\newcommand\dotdist{\pgflinewidth}}
    \fill (1*\dotdist,0) circle (0.5*\dotlw);
    \fill (2.5*\dotdist,0) circle (0.5*\dotlw);
    \fill (4*\dotdist,0) circle (0.5*\dotlw);
}}}}
\tikzset{startdots/.style= {postaction=decorate,decoration={name=markings,mark=at position 0 with{
   \ifdimless{\pgflinewidth}{1pt}{\newcommand\dotlw{1pt}}{\newcommand\dotlw{\pgflinewidth}}
   \ifdimless{\pgflinewidth}{1.5pt}{\newcommand\dotdist{1.5pt}}{\newcommand\dotdist{\pgflinewidth}}
   \fill (-1*\dotdist,0) circle (0.5*\dotlw);
   \fill (-2.5*\dotdist,0) circle (0.5*\dotlw);
   \fill (-4*\dotdist,0) circle (0.5*\dotlw);
}}}}

\tikzset{endlabel/.style={decoration={markings, mark=at position \pgfdecoratedpathlength-0.005cm with {\node at (0.2,0) {#1};}}, postaction={decorate}}}
\tikzset{startlabel/.style={decoration={markings, mark=at position 0.005cm with {\node at (-0.2,0) {#1};}}, postaction={decorate}}}
\tikzset{midlabeldistance/.store in=\midlabeldistance,midlabeldistance=0.2}
\tikzset{midlabelpos/.store in=\midlabelpos,midlabelpos=0.5}
\tikzset{midlabel/.style={decoration={markings, mark=at position \midlabelpos with {\node at (0,\midlabeldistance) {#1};}}, postaction={decorate}}}
\tikzset{midlabelr/.style={decoration={markings, mark=at position \midlabelpos with {\node at (0,-\midlabeldistance) {#1};}}, postaction={decorate}}}

\tikzset{forall/.style={decorate, decoration={snake,amplitude=0.04cm, segment length=0.1cm}}}
\tikzset{exists/.style={thick,dashed}}

\definecolor{ctlc0}{HTML}{000000}
\definecolor{ctlc1}{HTML}{61AAFF}
\definecolor{ctlc2}{HTML}{C74242}
\definecolor{ctlc3}{HTML}{E2C06C}
\definecolor{ctlc4}{HTML}{76E044}
\definecolor{ctlc5}{HTML}{DA7AD2}
\definecolor{ctlc6}{HTML}{FFA01F}
\definecolor{ctlc7}{HTML}{F41895}

\newcommand{\ifequals}[3]{\ifthenelse{\equal{#1}{#2}}{#3}{}}

\tikzset{backedge/.style={dashed}} 
\tikzset{mfat/.style={line width=1.3}} 
\tikzset{fat/.style={line width=2}} 
\tikzset{arrpos/.store in=\arrpos,arrpos=0.5}
\tikzset{arr/.style={postaction=decorate,decoration={name=markings,mark=at position \arrpos with{ 
    \draw[solid,line width=0.7] (#1 0.04,0)--(-#1 0.04,0.08) (#1 0.04,0)--(-#1 0.04,-0.08);
  }}}}
\tikzset{darrpos/.store in=\darrpos,darrpos=0.5}
\tikzset{darr/.style args={#1#2}{postaction=decorate,decoration={name=markings,mark=at position \darrpos with{ 
    \draw[solid] (0,0)--(#1 0.07,0)--(0,#2 0.07)--cycle;
  }}}}
\tikzset{darrbig/.style args={#1#2}{postaction=decorate,decoration={name=markings,mark=at position \darrpos with{ 
    \fill[black] (0,0)--(#1 0.1,0)--(0,#2 0.1)--cycle;
  }}}}
\tikzset{fatarrpos/.store in=\fatarrpos,fatarrpos=0.5}
\tikzset{fatarr/.style={postaction=decorate,decoration={name=markings,mark=at position \fatarrpos with{ 
    \draw[solid,line width=1] (#1 0.04,0)--(-#1 0.04,0.08) (#1 0.04,0)--(-#1 0.04,-0.08);
  }}}}
\tikzset{arrfat/.style={mfat,fatarr=#1}} 
\tikzset{hcircpos/.store in=\hcircpos,hcircpos=0.5}
\tikzset{hcirc/.style={postaction=decorate,decoration={name=markings,mark=at position \hcircpos with{ 
\draw (-0.1,0) arc (#1 180:0:0.1)--cycle;
}}}}

\tikzset{circarr/.style args={#1#2}{postaction=decorate,decoration={name=markings,mark=at position 0.5*\pgfdecoratedpathlength+0.15cm with{ 
\ifthenelse{\equal{#2}{}}{}{\draw (-0.1,0) arc (#2 180:0:0.1)--cycle;}
  }}, decoration={name=markings,mark=at position 0.5*\pgfdecoratedpathlength-0.15cm with{
\ifthenelse{\equal{#2}{}}{}{\draw[solid,line width=0.7] (#1 0.04,0)--(-#1 0.04,0.08) (#1 0.04,0)--(-#1 0.04,-0.08);}
  }}, decoration={name=markings,mark=at position 0.5 with{
\ifthenelse{\equal{#2}{}}{\draw[solid,line width=0.7] (#1 0.04,0)--(-#1 0.04,0.08) (#1 0.04,0)--(-#1 0.04,-0.08);}{}
  }}
}}
\tikzset{tcfedge/.style={}} 
\tikzset{overedge/.style={line width=3}} 
\tikzset{over/.style={transparency group, opacity=0.5}} 
\tikzset{overarr/.style={postaction=decorate,decoration={name=markings,mark=at position \arrpos with{ 
    \draw[solid,line width=2] (#1 0.08,0)--(-#1 0.08,0.16) (#1 0.08,0)--(-#1 0.08,-0.16);
  }}}}

\newcommand{\vertex}[2][black]{\fill[#1] #2 circle (0.075);}
\newcommand{\backvertex}[2][black]{\draw[#1,fill=white] #2 circle (0.075);}

\newcommand{\oververtex}[2][black]{\fill[#1] #2 circle (0.15);}
\newcommand{\edgetensor}[3][black]{\begin{scope}[shift={#3}] \draw[#1,rotate=#2,fill=white] (-0.15,-0.15) rectangle (0.15,0.15); \end{scope}}
\newcommand{\facetensor}[3][black]{\begin{scope}[shift={#3}] \draw[#1,rotate=#2,fill=white] (-30:0.2)--(90:0.2)--(-150:0.2)--cycle; \end{scope}}

\newcommand{\trianglebot}[4][black]{\begin{scope}[shift={#3}] \draw[#1,rotate=#2,fill=white] (-30:0.2)--(90:0.2)--(-150:0.2)--cycle; \fill[#1,rotate=#2] (-90 #4 60:0.2)--(0,-0.5*0.2)--(0,0)--cycle; \end{scope}}
\newcommand{\squarebot}[4][black]{\begin{scope}[shift={#3}] \draw[#1,rotate=#2,fill=white] (-0.15,-0.15) rectangle (0.15,0.15); \draw[#1,rotate=#2,line width=2.5] (-0.15,-0.15)--(0.15,-0.15); \end{scope}}
\newcommand{\squaremark}[4][black]{\begin{scope}[shift={#3}] \draw[#1,rotate=#2,fill=white] (-0.15,-0.15) rectangle (0.15,0.15); \fill[#1,rotate=#2] (#4 0.15,-0.15)--(0,-0.15)--(0,0)--cycle; \end{scope}}

\definecolor{markcol}{rgb}{1,0,0}
\tikzset{linemark/.style={postaction={draw,red,opacity=0.3,line width=0.15cm,densely dotted}}} 
\tikzset{areamark/.style={pattern=crosshatch,pattern color=red,opacity=0.5}} 
\tikzset{hatch/.style={pattern=crosshatch,pattern color=black}} 

\tikzset{facelab/.style={}}
\tikzset{backlabel/.style={text=gray}}
\tikzset{plabel/.style={font=\scriptsize}} 
\tikzset{ilabel/.style={font=\normalsize}} 

\tikzset{line/.style={}}
\tikzset{backline/.style={dash pattern={on 3pt off 1pt}}}
\tikzset{ctlf/.style={fill=#1,fill opacity=0.3,text opacity=1}}
\tikzset{ctldf/.style={fill=#1,fill opacity=0.51,text opacity=1}}

\tikzset{irrep/.style={line width=1.5}}
\tikzset{block/.style={}}
\tikzset{fusion/.style={decorate, decoration={snake,amplitude=0.04cm, segment length=0.1cm}}}
\tikzset{real/.style={}}
\tikzset{cstar/.style={}}
\tikzset{whstar/.style={}}
\tikzset{complex/.style={densely dotted}}
\tikzset{carrsoff/.store in=\carrsoff,carrsoff=0.3}
\tikzset{carreoff/.store in=\carreoff,carreoff=0.3}
\tikzset{carrs/.style={postaction=decorate,decoration={name=markings,mark=at position \carrsoff cm with{
    \draw[solid,semithick,sharp corners] (-#1 0.04,0.07)--(#1 0.04,0)--(-#1 0.04,-0.07);
  }}}}
\tikzset{carre/.style={postaction=decorate,decoration={name=markings,mark=at position \pgfdecoratedpathlength -\carreoff cm with{
    \draw[solid,semithick,sharp corners] (-#1 0.04,0.07)--(#1 0.04,0)--(-#1 0.04,-0.07);
  }}}}
\tikzset{realified/.style={line width=1.5}}
\tikzset{multiplicity/.style={decorate, decoration={snake,amplitude=0.04cm, segment length=0.1cm}}}
\tikzset{doubleindex/.style={thick}}

\tikzset{flags/.style={postaction=decorate,decoration={name=markings,mark=at position 0.26cm with{
    \draw[fill,solid,semithick,sharp corners] (-0.06,0)--(0,#1 0.09)--(0.06,0)--cycle;
  }}}}
\tikzset{flage/.style={postaction=decorate,decoration={name=markings,mark=at position \pgfdecoratedpathlength -0.26cm with{
    \draw[fill,solid,semithick,sharp corners] (-0.06,0)--(0,#1 0.09)--(0.06,0)--cycle;
  }}}}

\tikzset{tensorbox/.style={rectangle, draw, inner sep=3pt}}
\tikzset{tightbox/.style={rectangle, draw, inner sep=1pt}}
\newcommand{\circlenode}[2]{\node[circle,draw,fill=white,inner sep=0.015cm,minimum width=0.4cm] at #2 {#1};}

\newcommand{\trianglepattern}[4]{
\foreach \rel in {1,...,#4}{
\draw ($#1!\rel/#4!#2$)--($#1!\rel/#4!#3$);
\draw ($#2!\rel/#4!#1$)--($#2!\rel/#4!#3$);
\draw ($#3!\rel/#4!#1$)--($#3!\rel/#4!#2$);
}
}

\begin{document}
\title{Generalized topological state-sum constructions and their universality}
\author{A.~Bauer}
\affil{Dahlem Center for Complex Quantum Systems, Freie Universit{\"a}t Berlin,
Arnimallee 14, 14195 Berlin}

\maketitle

\begin{abstract}
We formalize and generalize the concept of a topological state-sum construction using the language of tensor networks. We give examples for constructions that are possibly more general than all state-sum constructions in the literature that we are aware of. In particular we propose a state-sum construction that is universal in the sense that it can emulate every other state-sum construction. Physically, state-sum models in $n$ dimensions correspond to fixed point models for topological phases of matter in $n$ space-time dimensions. We conjecture that our universal state-sum construction contains fixed point models for topological phases that are not captured by known constructions. In particular we demonstrate that, unlike common state-sum constructions, the construction is compatible with the absence of gapped boundaries and commuting-projector Hamiltonians in $2+1$-dimensional chiral topological phases.
\end{abstract}
\section{Introduction}
State-sum constructions are an important tool in many-body physics. On the one hand they have been used for generating fixed point models for topologically ordered phases in many-body physics. On the other hand they are concrete regularized representations of topological field theories in high energy physics.

A topological state-sum model can be seen as a prescription that associates tensor-networks to discretized versions (e.g. triangulations) of topological manifolds. The notion of locality in the tensor-network has to coincide with that of the triangulation, i.e. the tensors are associated to local components of the triangulation, and only indices of tensors that are near each other in the triangulation can be contracted. The topological invariance of the construction can then be formulated as a set of constraints on the tensors arising from local (Pachner) moves acting on the triangulation.

For state-sum constructions known from the literature, the prescription that distributes tensors and contractions over the triangulations is particularly simple: Usually one associates one and the same tensor to every simplex of the triangulation, and contracts indices of neighboring simplices. Examples are the constructions known as lattice TQFT, such as
\begin{itemize}
\item The models by Fukuma et al. \cite{Fukuma1992} for the $2$-dimensional case.
\item The Turaev-Viro state-sum \cite{Turaev1992,Barrett1993} (or its Hamiltonian formulation \cite{Levin2004}, or its PEPS formulation \cite{Bultinck2015}) in $3$ dimensions.
\item The Dijkgraaf-Witten model \cite{Dijkgraaf1990} (or its Hamiltonian formulation \cite{Hu2012}) in $3$ dimensions.
\item The Crane-Yetter model \cite{Crane1993} (or its Hamiltonian formulation \cite{Walker2011}) in $4$ dimensions.
\end{itemize}

Similar is the Kuperberg invariant \cite{Kuperberg1990} (or its Hamiltonian formulation \cite{Kitaev1997,Buerschaper2010,Chang2013}, or its PEPS formulation \cite{Schuch2010}) which can be formulated as a state-sum construction that associates tensors to the faces and edges of a $3$-manifold triangulation (or cellulation), such that pairs of adjacent edges and faces share a contracted index. Also this construction is simple in the sense that the tensor associated to a face/edge only depends on the number of adjacent edges/faces. In fact, after sufficient generalization and using the right technical details, this state-sum construction is equivalent to the Turaev-Viro state-sum \cite{tensor_lattice}.

In this paper we consider more general state-sum constructions where the tensor and contractions associated with a certain point in the triangulation is allowed to depend on a small (constant-size) environment of the triangulation around that point. We show that there are state-sum constructions which are universal in the following sense: They can emulate every other state-sum construction after sufficient coarse-graining and reshaping of the tensor network. Those universal state-sum constructions are however more complex than the state-sum constructions from the literature. In particular, in contrast to state-sum like Turaev-Viro, the canonical constructions for topological boundaries and commuting-projector Hamiltonians fail for universal state-sum constructions. Thus, they are compatible with chiral topological phases in $2+1$ dimensions, which by definition do not possess topological (gapped) boundaries, and are known not to allow for commuting-projector Hamiltonians. Universal state-sum constructions could potentially provide a systematic way to construct exactly solvable models for all (including chiral) topological phases, and a classification on a microscopic physical level as solutions of a finite set of polynomial equations.

The structure of the paper is as follows. In Sec.~(\ref{sec:tensor_lattices}) we formalize the concept of a state-sum construction as a so-called \emph{tensor lattice type} (short \emph{TL type}). In Sec.~(\ref{sec:universal_1d},\ref{sec:universal_2d}) we present concrete TL types in $1$ and $2$ dimensions, and prove their universality. In Sec.~(\ref{sec:universal_higherd}) we sketch how this construction can be generalized to arbitrary dimensions. In Sec.~(\ref{sec:top_boundary}) we describe how standard constructions for topological boundaries, commuting-projector Hamiltonians and tensor-network ground states fail for the generalized (universal) state-sum constructions, and how this is related to chiral phases. In Sec.~(\ref{sec:outlook}) we give ideas how to construct instances of the generalized state-sum constructions describing phases of matter that are lacking a fixed point description so far.

\section{Topological tensor lattices}
\label{sec:tensor_lattices}
\subsection{Topological lattices}
\label{sec:lattices}
Intuitively, a \emph{$n$-dimensional simplicial complex} is a decomposition of a topological $n$-manifold into $n$-simplices that share common $(n-1)$-simplices at their boundary. Those $(n-1)$-simplices in turn meet at $(n-2)$-simplices and so on. $0$-, $1$-, $2$-, and $3$-simplices are referred to as \emph{vertices}, \emph{edges}, \emph{triangles} and \emph{tetrahedra}. Additionally, a $n$-dimensional simplicial complex will usually include an orientation of its edges that is never cyclic around any face, commonly referred to as \emph{branching structure}. In some cases we don't include a branching structure, or have some other kind of decoration instead. Consider the following examples of simplicial complexes:
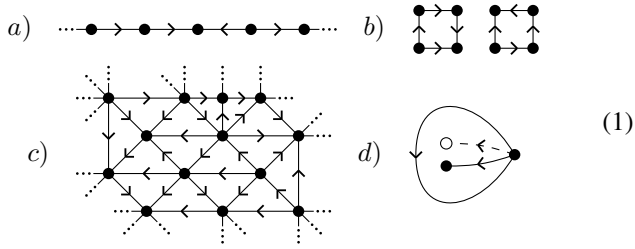
\begin{equation}
\begin{gathered}
\begin{aligned}
a)\quad &
\begin{tikzpicture}
\coords{0/0,0|1/0.7,0|2/1.4,0|3/2.1,0|4/2.8,0};
\draw (0)edge[arr=+](1) (1)edge[arr=+](2) (3)edge[arr=+](2) (3)edge[arr=+](4);
\draw (0)edge[enddots]++(-0.2,0) (4)edge[enddots]++(0.2,0);
\shapes{\vertex}{0,1,2,3,4};
\end{tikzpicture}&
b)\quad &
\begin{tikzpicture}
\coords{0/0,0|1/0.5,0|2/0,0.5|3/0.5,0.5|4/1,0|5/1.5,0|6/1,0.5|7/1.5,0.5};
\draw (0)edge[arr=+](1) (0)edge[arr=+](2)(3)edge[arr=+](1)(2)edge[arr=+](3)  (4)edge[arr=+](5) (4)edge[arr=+](6)(7)edge[arr=-](5)(6)edge[arr=-](7);
\shapes{\vertex}{0,1,2,3,4,5,6,7};
\end{tikzpicture}
\end{aligned}\\
\begin{aligned}
c)&\quad
\begin{tikzpicture}
\coords{0/0,0|1/1,0|2/0.5,0.5|3/-0.5,0.5|4/-1,0|5/-0.5,-0.5|6/0.5,-0.5|7/1.5,0.5|8/0,1|9/0.5,1|10/1,1|11/1.5,-0.5|12/-1,1};
\draw (0)edge[arr=-](1) (2)edge[arr=+](0) (0)edge[arr=+](3) (0)edge[arr=+](4) (0)edge[arr=+](5) (0)edge[arr=+](6) (1)edge[arr=+](6) (6)edge[arr=+](5) (4)edge[arr=+](5) (3)edge[arr=+](4) (2)edge[arr=+](3) (1)edge[arr=+](2) (2)edge[arr=+](7) (2)edge[arr=+](10) (2)edge[arr=+](9) (8)edge[arr=+](2) (8)edge[arr=+](9) (9)edge[arr=+](10) (10)edge[arr=+](7) (7)edge[arr=-](1) (11)edge[arr=+](7) (11)edge[arr=+](1) (11)edge[arr=+](6) (8)edge[arr=+](3) (12)edge[arr=+](3) (4)edge[arr=-](12) (12)edge[arr=+](8);
\draw (11)edge[enddots]++(-90:0.2) (11)edge[enddots]++(-45:0.2) (11)edge[enddots]++(0:0.2) (7)edge[enddots]++(0:0.2) (7)edge[enddots]++(45:0.2) (10)edge[enddots]++(0:0.2) (10)edge[enddots]++(90:0.2) (9)edge[enddots]++(90:0.2) (8)edge[enddots]++(90:0.2) (8)edge[enddots]++(135:0.2) (12)edge[enddots]++(90:0.2) (12)edge[enddots]++(135:0.2) (12)edge[enddots]++(180:0.2) (12)edge[enddots]++(-135:0.2) (4)edge[enddots]++(180:0.2) (4)edge[enddots]++(-135:0.2) (5)edge[enddots]++(180:0.2) (5)edge[enddots]++(-135:0.2) (5)edge[enddots]++(-45:0.2) (6)edge[enddots]++(-90:0.2) (6)edge[enddots]++(-45:0.2);
\shapes{\vertex}{0,1,2,3,4,5,6,7,8,9,10,11,12}:
\end{tikzpicture}&
d)\quad &
\begin{tikzpicture}
\coords{0/0,0|1/-0.9,-0.15|2/-0.9,0.15};
\draw[arr=+] (0)to[out=120,in=90,looseness=1.8]++(-1.3,0)to[out=-90,in=-120,looseness=1.8](0);
\draw (0)edge[arr=+,bend left=10](1) (0)edge[backedge,arr=+,bend right=10](2);
\shapes{\vertex}{0,1}
\backvertex{(2)};
\end{tikzpicture}
\end{aligned}
\end{gathered}
\end{equation}
a) shows a patch of a $1$-dimensional simplicial complex. b) shows a complete $1$-dimensional simplicial complex on a disjoint union of two circles. c) shows a patch of a $2$-dimensional simplicial complex. d) shows a complete $2$-dimensional simplicial complex representing a sphere (each with one vertex, one edge and one triangle in the front and the back layer, and another vertex and edge where the two layers meet).

A state-sum construction only depends on the combinatorics of the simplicial complex, i.e. which $(n-1)$-simplices are shared by which $n$-simplices in which way. This already encodes all topological properties of the manifold. The geometric details of the decomposition are not important. We will refer to this combinatorial data as \emph{$n$tS lattices} (``t'' for ``topological'' and ``S'' for ``simplicial''). In \cite{tensor_lattice} we made this purely combinatorial viewpoint explicit.

A $x$ to $n+2-x$ \emph{Pachner move} is the following prescription that changes a $n$tS lattice locally: The boundary of a $(n+1)$-simplex forms a $n$tS lattice with $n+2$ $n$-simplices. Consider a bipartition of this $n$tS lattice into two parts with $x$ and $n+2-x$ $n$-simplices. The corresponding Pachner move takes a patch of the complex that looks like the one part and replaces it with the other part. Consider the following examples of Pachner moves:
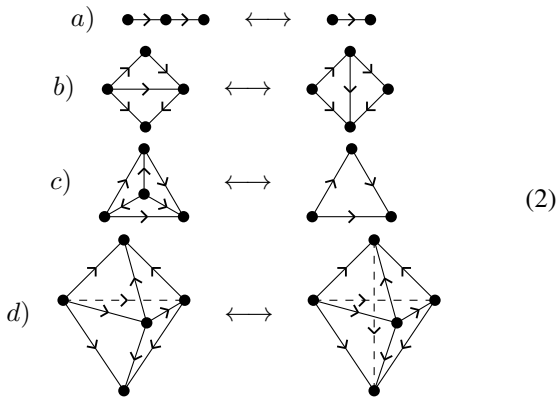
\begin{equation}
\begin{gathered}
a)\quad
\begin{tikzpicture}
\coords{0/0,0|1/0.5,0|2/1,0};
\draw (0)edge[arr=+](1) (1)edge[arr=+](2);
\shapes{\vertex}{0,1,2};
\end{tikzpicture}\quad\longleftrightarrow\quad
\begin{tikzpicture}
\coords{0/0,0|1/0.5,0};
\draw (0)edge[arr=+](1);
\shapes{\vertex}{0,1};
\end{tikzpicture}\\
b)\quad
\begin{tikzpicture}
\coords{0/0,0|1/0.5,-0.5|2/1,0|3/0.5,0.5};
\draw (0)edge[arr=+](3) (3)edge[arr=+](2) (2)edge[arr=+](1) (0)edge[arr=+](1) (0)edge[arr=+](2);
\shapes{\vertex}{0,1,2,3};
\end{tikzpicture}\quad \longleftrightarrow\quad
\begin{tikzpicture}
\coords{0/0,0|1/0.5,-0.5|2/1,0|3/0.5,0.5};
\draw (0)edge[arr=+](3) (3)edge[arr=+](2) (2)edge[arr=+](1) (0)edge[arr=+](1) (3)edge[arr=+](1);
\shapes{\vertex}{0,1,2,3};
\end{tikzpicture}\\
c)\quad
\begin{tikzpicture}
\coords{0/0,0|1/-30:0.6|2/90:0.6|3/-150:0.6};
\draw (0)edge[arr=+](1) (0)edge[arr=+](2) (0)edge[arr=+](3) (3)edge[arr=+](2) (2)edge[arr=+](1) (3)edge[arr=+](1);
\shapes{\vertex}{0,1,2,3};
\end{tikzpicture} \quad\longleftrightarrow\quad
\begin{tikzpicture}
\coords{1/-30:0.6|2/90:0.6|3/-150:0.6};
\draw (3)edge[arr=+](2) (2)edge[arr=+](1) (3)edge[arr=+](1);
\shapes{\vertex}{1,2,3};
\end{tikzpicture}\\
d)\quad
\begin{tikzpicture}
\coords{0/0,0|1/1.6,0|2/1.1,-0.3|3/0.8,0.8|4/0.8,-1.2};
\draw (0)edge[backedge,circarr=+](1) (0)edge[circarr=+](2) (2)edge[circarr=+](1) (0)edge[circarr=+](3) (1)edge[circarr=+](3) (2)edge[circarr=+](3) (0)edge[circarr=+](4) (1)edge[circarr=+](4) (2)edge[circarr=+](4);
\shapes{\vertex}{0,1,2,3,4};
\end{tikzpicture}\quad\longleftrightarrow\quad
\begin{tikzpicture}
\coords{0/0,0|1/1.6,0|2/1.1,-0.3|3/0.8,0.8|4/0.8,-1.2};
\draw (0)edge[backedge,arr=+,arrpos=0.4](1) (0)edge[circarr=+](2) (2)edge[circarr=+](1) (0)edge[circarr=+](3) (1)edge[circarr=+](3) (2)edge[circarr=+](3) (0)edge[circarr=+](4) (1)edge[circarr=+](4) (2)edge[circarr=+](4) (3)edge[backedge,arr=+,arrpos=0.6](4);
\shapes{\vertex}{0,1,2,3,4};
\end{tikzpicture}
\end{gathered}
\end{equation}
a) shows a $2$ to $1$ Pachner move acting on a patch of a $1$tS lattice, b) and c) show a $2$ to $2$ Pachner move and a $3$ to $1$ Pachner move acting on a patch of a $2$tS lattice, and d) shows a $2$ to $3$ Pachner move acting on a patch of a $3$tS lattice. The left side shows two tetrahedra separated by one horizontal face, whereas the right side consists of three tetrahedra, meeting at one vertical edge in the middle, and separated by three vertical faces.

It is known that two triangulations of topological (more precisely, piece-wise linear) manifolds can be deformed into each other via Pachner moves iff the manifolds are homeomorphic. So there's a one-to-one relation between equivalence classes of $n$tS lattices under Pachner moves and equivalence classes of topological manifolds under homeomorphism. In this sense $n$tS lattices are a discrete combinatorial version of topological manifolds.

A \emph{lattice mapping} is a local prescription for a global transformation on $n$tS lattices. More precisely, such a prescription translates certain local configurations of the source $n$tS lattice into simplices of the target $n$tS lattice. We will depict such lattice mappings by drawing a representative lattice patch and the result of the mapping for that patch on top of it. For the original lattice patch we will use thicker lines and shapes and a semi-transparent and different color. E.g. consider the following lattice mappings:
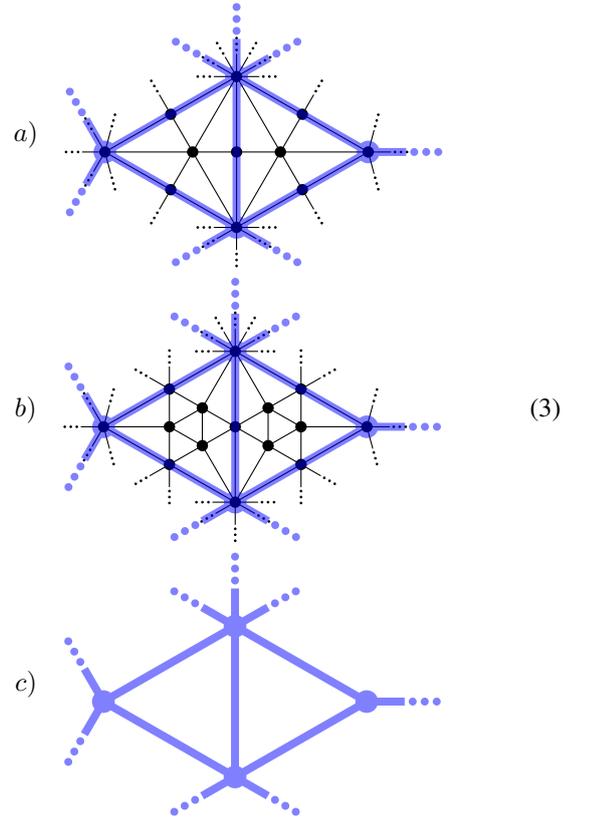
\begin{equation}
\label{eq:lattice_mapping_examples}
\begin{gathered}
a)\quad
\begin{tikzpicture}
\coords{0/0,0|1/30:2|2/150:2|3/0,2};
\coordinate (e0) at ($(0)!0.5!(1)$);
\coordinate (e1) at ($(0)!0.5!(2)$);
\coordinate (e2) at ($(0)!0.5!(3)$);
\coordinate (e3) at ($(1)!0.5!(3)$);
\coordinate (e4) at ($(2)!0.5!(3)$);
\coordinate (f0) at ({sqrt(3)/3},1);
\coordinate (f1) at ({-sqrt(3)/3},1);
\draw (0)--(e0)--(1)--(e3)--(3)--(e4)--(2)--(e1)--(0) (0)--(e2)--(3) (f0)--(e0) (f0)--(e2) (f0)--(e3) (f1)--(e1) (f1)--(e2) (f1)--(e4) (f0)--(0) (f0)--(3) (f1)--(0) (f1)--(3) (f1)--(2) (f0)--(1);
\draw (0)edge[enddots]++(0:0.3) (0)edge[enddots]++(180:0.3) (0)edge[enddots]++(-90:0.3) (0)edge[enddots]++(-30:0.3) (0)edge[enddots]++(-150:0.3) (1)edge[enddots]++(0:0.3) (1)edge[enddots]++(75:0.3) (1)edge[enddots]++(-75:0.3) (2)edge[enddots]++(-120:0.3) (2)edge[enddots]++(120:0.3) (2)edge[enddots]++(180:0.3) (2)edge[enddots]++(75:0.3) (2)edge[enddots]++(-75:0.3) (3)edge[enddots]++(30:0.3) (3)edge[enddots]++(60:0.3) (3)edge[enddots]++(90:0.3) (3)edge[enddots]++(120:0.3) (3)edge[enddots]++(150:0.3) (3)edge[enddots]++(0:0.3) (3)edge[enddots]++(180:0.3) (e0)edge[enddots]++(-60:0.3) (e1)edge[enddots]++(-120:0.3) (e3)edge[enddots]++(60:0.3) (e4)edge[enddots]++(120:0.3);
\shapes{\vertex}{e0,e1,e2,e3,e4,f0,f1,0,1,2,3};
\begin{scope}[over]
\draw[overedge,blue] (0)--(1)--(3)--(2)--(0) (0)--(3) (0)edge[enddots]++(-30:0.5) (0)edge[enddots]++(-150:0.5) (1)edge[enddots]++(0:0.5) (2)edge[enddots]++(-120:0.5) (2)edge[enddots]++(120:0.5) (3)edge[enddots]++(30:0.5) (3)edge[enddots]++(90:0.5) (3)edge[enddots]++(150:0.5);
\shapes{\oververtex[blue]}{0,1,2,3};
\end{scope}
\end{tikzpicture}\\
b)\quad
\begin{tikzpicture}
\coords{0/0,0|1/30:2|2/150:2|3/0,2};
\coordinate (e0) at ($(0)!0.5!(1)$);
\coordinate (e2) at ($(0)!0.5!(2)$);
\coordinate (e1) at ($(0)!0.5!(3)$);
\coordinate (e3) at ($(1)!0.5!(3)$);
\coordinate (e4) at ($(2)!0.5!(3)$);
\coordinate (e01) at ($(e0)!0.5!(e1)$);
\coordinate (e12) at ($(e1)!0.5!(e2)$);
\coordinate (e03) at ($(e0)!0.5!(e3)$);
\coordinate (e13) at ($(e1)!0.5!(e3)$);
\coordinate (e24) at ($(e2)!0.5!(e4)$);
\coordinate (e14) at ($(e1)!0.5!(e4)$);
\draw (0)--(e0)--(1)--(e3)--(3)--(e4)--(2)--(e2)--(0) (0)--(e1)--(3) (e0)--(e01)--(e1) (e1)--(e12)--(e2) (e0)--(e03)--(e3) (e1)--(e13)--(e3) (e2)--(e24)--(e4) (e1)--(e14)--(e4) (e01)--(e13)--(e03)--cycle (e12)--(e24)--(e14)--cycle (0)--(e01) (0)--(e12) (1)--(e03) (2)--(e24) (3)--(e13) (3)--(e14);
\draw (0)edge[enddots]++(0:0.3) (0)edge[enddots]++(180:0.3) (0)edge[enddots]++(-90:0.3) (0)edge[enddots]++(-30:0.3) (0)edge[enddots]++(-150:0.3) (1)edge[enddots]++(0:0.3) (1)edge[enddots]++(75:0.3) (1)edge[enddots]++(-75:0.3) (2)edge[enddots]++(-120:0.3) (2)edge[enddots]++(120:0.3) (2)edge[enddots]++(180:0.3) (2)edge[enddots]++(75:0.3) (2)edge[enddots]++(-75:0.3) (3)edge[enddots]++(30:0.3) (3)edge[enddots]++(60:0.3) (3)edge[enddots]++(90:0.3) (3)edge[enddots]++(120:0.3) (3)edge[enddots]++(150:0.3) (3)edge[enddots]++(0:0.3) (3)edge[enddots]++(180:0.3) (e0)edge[enddots]++(-90:0.3) (e2)edge[enddots]++(-150:0.3) (e3)edge[enddots]++(30:0.3) (e4)edge[enddots]++(90:0.3) (e0)edge[enddots]++(-30:0.3) (e2)edge[enddots]++(-90:0.3) (e3)edge[enddots]++(90:0.3) (e4)edge[enddots]++(150:0.3);
\shapes{\vertex}{e0,e1,e2,e3,e4,e01,e12,e03,e13,e24,e14,0,1,2,3};
\begin{scope}[over]
\draw[overedge,blue] (0)--(1)--(3)--(2)--(0) (0)--(3) (0)edge[enddots]++(-30:0.5) (0)edge[enddots]++(-150:0.5) (1)edge[enddots]++(0:0.5) (2)edge[enddots]++(-120:0.5) (2)edge[enddots]++(120:0.5) (3)edge[enddots]++(30:0.5) (3)edge[enddots]++(90:0.5) (3)edge[enddots]++(150:0.5);
\shapes{\oververtex[blue]}{0,1,2,3};
\end{scope}
\end{tikzpicture}\\
c)\quad
\begin{tikzpicture}
\coords{0/0,0|1/30:2|2/150:2|3/0,2};
\begin{scope}[over]
\draw[overedge,blue] (0)--(1)--(3)--(2)--(0) (0)--(3) (0)edge[enddots]++(-30:0.5) (0)edge[enddots]++(-150:0.5) (1)edge[enddots]++(0:0.5) (2)edge[enddots]++(-120:0.5) (2)edge[enddots]++(120:0.5) (3)edge[enddots]++(30:0.5) (3)edge[enddots]++(90:0.5) (3)edge[enddots]++(150:0.5);
\shapes{\oververtex[blue]}{0,1,2,3};
\end{scope}
\end{tikzpicture}
\end{gathered}
\end{equation}
a) shows a lattice mapping known as the \emph{barycentric subdivision}. b) shows a fine-graining mapping dividing each edge into two edges and each triangle into $10$ triangles. c) shows the \emph{trivial lattice mapping} which maps every lattice to the empty one.

Lattice mappings can change the topology, for example by attaching small handles everywhere. Another example is the trivial lattice mapping above. For the sake of this paper we are mostly interested in mappings that do not change the topology (i.e. that can be performed by a circuit of moves), such as the mappings a) and b) in Eq.~(\ref{eq:lattice_mapping_examples}). Such mappings can be thought of as fine-grainings, as they replace each simplex by an arbitrarily fine-grained patch of simplices. Conversely, they cannot preform coarse-graining in the sense of blocking multiple unit cells into one, as there is no canonical way to locally determine where the coarser unit cell should start and end.

\subsubsection*{Lattice types and classes}
Note that there are many alternative combinatorial formulations of topological manifolds, such as general cell complexes. In \cite{tensor_lattice} we formulated a general axiomatic framework for defining data structures with a notion of locality and moves, that we called \emph{lattice types}. Locality is enforced via the \emph{local finiteness principle}, asserting that the number of possible lattice configurations within a patch of fixed size (i.e. diameter measured in the combinatorial lattice distance) must be finite.

Lattice mappings can be defined more generally as local prescriptions mapping between lattices of different types. The different lattice types can be divided into \emph{classes} which are equivalent up to invertible lattice mappings. Types in the same class roughly correspond to different combinatorial representations of the same emergent continuum structure. ``Topological'' lattice types are a class whose emergent continuum structure is topological manifolds (more precisely, for each dimension $n$ there is a separate class). We can without loss of generality restrict to $n$tS lattices as one particular lattice type in the corresponding class.

There is one subtlety concerning $n$tS lattices: In a $n$-dimensional simplicial complex, the number of $n$-simplices around a vertex can be arbitrary. So there are infinitely many configurations for the $n$-simplices around the vertex, violating the local finiteness principle. There are two possible solutions to this problem:

1) The combinatorial data includes the $n$-simplices and the information about which $(n-1)$-simplices they share, however the vertices are not explicitly part of it. Now consider a vertex with a large number of surrounding $n$-simplices and two of the $n$-simplices on opposite sides of the vertex. Even though they share a common vertex, those $n$-simplices are actually far apart when we measure in the combinatorial lattice distance, as we have to go from one to the other via all the $n$-simplices and $(n-1)$-simplices in between. Conversely, if we fix a size of a lattice patch (in the combinatorial distance), a patch of this size cannot contain all $n$-simplices around a vertex for every configuration of $n$-simplices around a vertex, so there is no violation of the local finiteness principle.

2) We can explicitly forbid more than $l$ $n$-simplices around a vertex for some $l$. To do this in a more controlled way, one can define the \emph{link} of a $x$-simplex as the $(n-x-1)$tS lattice formed by all the $(n-x-1)$-simplices that together with the $x$-simplex span a $n$-simplex. Instead of restricting the bare number of $n$-simplices around a vertex, we can restrict to a finite set of allowed links for $x$-simplices, for all different $x$. This set of allowed links (or the number $l$) has to be large enough such that we can still represent any topological manifold and transform any two triangulations of the same manifold into each other via Pachner moves. E.g. $l=5$ in $2$ dimensions (i.e. no more than $5$ triangles around every vertex) is not sufficient. With this limitation we get only a finite number of lattices, the largest one being an icosahedron.

The two approaches yield two different lattice types in different classes. We will let the name ``$n$tS lattices'' refer to the types 2) and will call lattices of the type 1) \emph{$n$taS lattices} (where ``a'' stands for ``arbitrary link''). The different $n$tS types for different sets of allowed links (or different numbers $l$) are all in the same lattice class as long as the sets (number $l$) are (is) large enough. Every $n$tS lattice can be interpreted as a $n$taS lattice in the obvious way, defining a lattice mapping from $n$tS lattices to $n$taS lattices.

Both $n$tS and $n$taS have topological $n$-manifolds as their continuum picture. However, the latter are better represented by $n$tS: For $n$taS lattices, points on different $n$-simplices can be close on the $n$-manifold whereas the two $n$-simplices are far apart measured in the combinatorial lattice distance. So only $n$tS should be regarded as a proper ``topological'' type.

\subsection{Real tensors and tensor networks}
In this section we will quickly recap real tensors and how they are denoted in \emph{tensor-network notation}.

A \emph{real tensor} is defined with respect to a finite set of indices $i\in I$ each of which is equipped with a finite set $B_i$ called \emph{basis}. It is a map $T$ that associates a real number to every element of the cartesian product of all the $B_i$:
\begin{equation}
T:\quad \bigtimes_{i\in I} B_i \rightarrow \mathbb{R}
\end{equation}
To avoid confusion we should add the following remarks: 1) We are not interested to hypothetical transformation properties of a tensor which is often understood as part of the definition in differential geometry. For us a tensor is just a collection of real numbers. 2) We are slightly abusing the word ``basis'': For us this does not refer to a subset of vectors in a vector space but simply to a finite set.

In tensor-network notation a real tensor is denoted by a labeled box or any other shape with lines emanating from the boundary whose endpoints carry labels. Each of those lines corresponds to one index. E.g., a $3$-index tensor $T$ could be denoted as:
\begin{equation}
\begin{gathered}
a)\quad
\begin{tikzpicture}
\node[draw,rectangle] (t) {$T$};
\draw (t.west)--++(-0.5,0) node[left]{$a$};
\draw (t.north)--++(0,0.3) node[above]{$b$};
\draw (t.east)--++(0.5,0) node[right]{$c$};
\end{tikzpicture}\qquad b)\quad
\begin{tikzpicture}
\node[draw,circle] (t) {$T$};
\draw (t.west)--++(-0.5,0) node[left]{$a$};
\draw (t.north)--++(0,0.3) node[above]{$b$};
\draw (t.east)--++(0.5,0) node[right]{$c$};
\end{tikzpicture}\\
c)\quad
\begin{tikzpicture}
\coords{0/0,0};
\draw (0)--++(-0.6,0) node[left]{$a$};
\draw (0)--++(0,0.4) node[above]{$b$};
\draw (0)--++(0.6,0) node[right]{$c$};
\begin{scope}
\draw[thick,fill=white] (0)circle(0.2);
\clip (0)circle(0.2);
\draw[thick] (-0.2,0.2)--(0.2,-0.2) (-0.2,-0.2)--(0.2,0.2);
\end{scope}
\end{tikzpicture}
\end{gathered}
\end{equation}
The labels at the line endings tell us how to match up indices when we equate two tensors. Often we will omit the labels and instead use the position of the endpoints and/or the directions they are pointing at to indicate the latter. Which index is which is determined by at which point the corresponding line is emanating from the shape/box. If a tensor occurs multiple times, the corresponding shape/box might be drawn in a rotated or reflected manner for the different occurrences. If a shape/box has rotation or reflection symmetries, the tensors are assumed to have the same symmetries acting by index permutation. E.g., if the shape is round, this means the tensors are invariant under index permutations which are cyclic or reversing the order.

The \emph{tensor product} of two real tensors $T_1$ and $T_2$ is a tensor $T$ whose index set is the disjoint union of the index sets of $T_1$ and $T_2$. Its entries are products of the entries of $T_1$ and $T_2$:
\begin{equation}
\begin{gathered}
T(b^{(1)}_1,b^{(2)}_1,\ldots,b^{(1)}_2,b^{(2)}_2,\ldots)=\\
T_1(b^{(1)}_1,b^{(2)}_1,\ldots)\cdot T_2(b^{(1)}_2,b^{(2)}_2,\ldots)
\end{gathered}
\end{equation}

The tensor product is denoted by placing two tensors next to each other, e.g., for a $3$-index tensor $T_1$ and a $2$-index tensor $T_2$ we get:
\begin{equation}
\begin{tikzpicture}
\node[draw,rectangle] (t) {$T$};
\draw (t.west)--++(-0.5,0) node[left]{$a$};
\draw (t.north west)--++(-0.5,0.3) node[left]{$d$};
\draw (t.north east)--++(0.5,0.3) node[right]{$e$};
\draw (t.north)--++(0,0.3) node[above]{$b$};
\draw (t.east)--++(0.5,0) node[right]{$c$};
\end{tikzpicture}=
\begin{tikzpicture}
\node[draw,rectangle] (t) {$T_1$};
\draw (t.west)--++(-0.5,0) node[left]{$a$};
\draw (t.north)--++(0,0.3) node[above]{$b$};
\draw (t.east)--++(0.5,0) node[right]{$c$};
\end{tikzpicture}
\begin{tikzpicture}
\node[draw,rectangle] (t) {$T_2$};
\draw (t.west)--++(-0.5,0) node[left]{$d$};
\draw (t.east)--++(0.5,0) node[right]{$e$};
\end{tikzpicture}
\end{equation}

Two indices of a tensor $T$ can be \emph{contracted} if they have the same basis $B$, yielding a tensor $T'$ with those two indices missing. $T'$ is obtained by setting the value of the two indices equal and summing over all different values:
\begin{equation}
\begin{gathered}
T'(b_1,\ldots,b_n)=\\\sum_{x\in B} T(\ldots,b_{a-1},x,b_{a+1},\ldots, b_{b-1},x,b_{b+1},\ldots)
\end{gathered}
\end{equation}

In tensor-network notation the contraction of two indices is denoted by connecting the corresponding lines. E.g., the contraction of the left and lower index of a $4$-index tensor $T$ is denoted as:
\begin{equation}
\begin{tikzpicture}
\node[draw,rectangle] (t) {$T'$};
\draw (t.west)--++(-0.5,0) node[left]{$a$};
\draw (t.east)--++(0.5,0) node[right]{$b$};
\end{tikzpicture}=
\begin{tikzpicture}
\node[draw,rectangle] (t) {$T$};
\draw[rounded corners] (t.west)--++(-0.5,0)--++(0,-0.5)-|(t.south);
\draw (t.north)--++(0,0.3) node[above]{$a$};
\draw (t.east)--++(0.5,0) node[right]{$b$};
\end{tikzpicture}
\end{equation}

A \emph{tensor network} is a set of tensors together with a prescription of how their indices are mutually contracted. The \emph{geometry} of the tensor network refers to the combinatorics of how many tensors there are, what their shapes are, and how their indices are contracted. The \emph{evaluation} of a tensor network refers to the tensor obtained by taking the tensor product of all the tensors and then carrying out all the contractions.

In tensor-network notation, tensor networks are represented by graph-like pictures. E.g.,
\begin{equation}
\begin{tikzpicture}
\node[draw,rectangle] (0) at (0,0) {$F$};
\node[draw,rectangle] (1) at (0,1) {$G$};
\node[draw,rectangle] (2) at (1,0) {$H$};
\node[draw,rectangle] (3) at (1,1) {$I$};
\draw (0.east)--(2.west) (1.east)--(3.west) (0.north)--(1.south) (2.north)--(3.south) (0.north east)--(3.south west) (1.south east)--(2.north west);
\draw (0.west)--++(-0.3,0)node[left]{$a$} (2.east)--++(0.3,0)node[right]{$b$} (1.west)--++(-0.3,0)node[left]{$c$} (3.east)--++(0.3,0)node[right]{$d$};
\end{tikzpicture}
\end{equation}
Two lines crossing does not have any effect.

The \emph{identity tensor} is consistently denoted by a free line, where the two endpoints of the line represent the two indices:
\begin{equation}
\begin{tikzpicture}
\draw (0,0)edge[startlabel=$a$,endlabel=$b$](1,0);
\end{tikzpicture}=
\delta_{a,b}=
\begin{cases}
1 \text{ if }a=b\\0 \text{ otherwise}
\end{cases}
\end{equation}

Real tensors are one specific \emph{tensor type}. There are other data structures with a notion of tensor product and contraction, yielding other tensor types from which we can also form tensor networks \cite {tensor_lattice, tensor_type}.

One might think that at least in quantum mechanics the use of complex numbers is more appropriate than real numbers. One can accordingly define complex tensors, which are completely analogous with complex instead of real entries. However, complex tensors can be simulated by real tensors via \emph{realification}. For the state-sum constructions, this has the additional advantages that one can get rid of both the orientation dependence and the unitarity condition \cite{tensor_lattice}.

\subsection{Topological tensor lattices}
A (real, $n$tS) \emph{tensor lattice} (short \emph{TL}) is a local prescription that associates real tensor-networks to $n$tS lattices, such that the axiom Eq.~(\ref{eq:tl_axiom}) below holds. By ``local prescription'' we mean that what the tensor network at one point looks like only depends on what the lattice looks like in a surrounding patch of constant size $L$. More precisely, $L$ is the maximal combinatorial lattice distance of two lattice components of the surrounding patch, and will be be referred to as the \emph{locality length} of the tensor lattice. The precise quantitative value of $L$ depends on the combinatorial details of the description and is therefore somewhat arbitrary. We are only interested in its scaling behavior, such as whether it stays constant or unbounded for a sequence of different TLs.

When two $n$tS lattices $L_a$ and $L_b$ differ locally by a Pachner move, the tensor-networks associated to the lattices also differ locally. Now consider patches $P_a$ and $P_b$ of size $L$ of the tensor network associated to $L_a$ and $L_b$ at the place where the move happens, such that the tensor networks of $L_a$ and $L_b$ are equal outside of those patches. $P_a$ and $P_b$ have to evaluate to the same tensor. The following diagram illustrates the situation:
\begin{equation}
\label{eq:tl_axiom}
\begin{tikzpicture}
\node (a) at (0,0) {$L_a$};
\node (b) at (4,0) {$L_b$};
\node (ta) at (0,-1.5) {$P_a$};
\node (tb) at (4,-1.5) {$P_b$};
\node (e) at (2, -3) {$T$};
\draw[->] (a.east)--node[midway,above]{Pachner move}(b.west);
\draw[->] (a.south)--node[midway,left]{patch}(ta.north);
\draw[->] (b.south)--node[midway,right]{patch}(tb.north);
\draw[->] (ta.south east)--node[midway,below left]{evaluate}(e.north west);
\draw[->] (tb.south west)--node[midway,below right]{evaluate}(e.north east);
\end{tikzpicture}
\end{equation}

The \emph{type} of the TL refers to the geometry of the tensor networks associated to the lattices, i.e. to which places the tensors are associated and in which way they are contracted. We refer to the precise choice of the patches $P_a$ and $P_b$ in Eq.~(\ref{eq:tl_axiom}) as the \emph{tensor-network moves} of the TL type.

We will now give a few examples in $2$ dimensions in order to demonstrate what possible TL types can look like. In the following we will depict TL types by drawing a representative lattice patch and drawing the associated tensor-network patch on top of it. Just as for lattice mappings the lattice components will be drawn fat and semi-transparent. The different tensors will be distinguished by using different shapes.

The standard example is to associate the same $3$-index tensor to every face and contract two copies of the tensors at faces sharing an edge. E.g.:
\begin{equation}
\label{eq:tl_example1}
\begin{tikzpicture}
\coords{0/0,0|1/30:2|2/150:2|3/0,2};
\coordinate (f0) at ({sqrt(3)/3},1);
\coordinate (f1) at ({-sqrt(3)/3},1);
\draw (f0)--(f1) (f1)edge[enddots]++(120:0.9) (f1)edge[enddots]++(-120:0.9) (f0)edge[enddots]++(60:0.9) (f0)edge[enddots]++(-60:0.9);
\trianglebot{-90}{(f0)}{-};
\trianglebot{-30}{(f1)}{-};
\begin{scope}[over]
\draw[overedge,blue] (0)edge[overarr=+](1)(1)edge[overarr=+](3)(3)edge[overarr=+](2)(2)edge[overarr=-](0)(0) (0)edge[overarr=+](3) (0)edge[enddots]++(-30:0.5) (0)edge[enddots]++(-150:0.5) (1)edge[enddots]++(-90:0.5) (1)edge[enddots]++(90:0.5) (2)edge[enddots]++(-90:0.5) (2)edge[enddots]++(90:0.5) (3)edge[enddots]++(30:0.5) (3)edge[enddots]++(150:0.5);
\shapes{\oververtex[blue]}{0,1,2,3};
\end{scope}
\end{tikzpicture}
\end{equation}
The canonical choice of tensor-network move are the tensors associated to the triangles directly involved in the Pachner move. E.g. for a $2$ to $2$ Pachner move we get the following equation:
\begin{equation}
\label{eq:pmove_sample}
\begin{tikzpicture}
\coords{0/0,0|1/0.5,-0.5|2/1,0|3/0.5,0.5};
\draw (0)edge[arr=-](3) (3)edge[arr=-](2) (2)edge[arr=-](1) (0)edge[arr=-](1) (3)edge[arr=-](1);
\shapes{\vertex}{0,1,2,3};
\end{tikzpicture}
\quad \longleftrightarrow\quad
\begin{tikzpicture}
\coords{0/0,0|1/0.5,-0.5|2/1,0|3/0.5,0.5};
\draw (0)edge[arr=-](3) (3)edge[arr=-](2) (2)edge[arr=-](1) (0)edge[arr=-](1) (0)edge[arr=-](2);
\shapes{\vertex}{0,1,2,3};
\end{tikzpicture}
\end{equation}
\begin{equation}
\begin{tikzpicture}
\coords{0/0,0|1/0.5,0};
\draw (0)--(1) (1)edge[endlabel=$a$]++(60:0.4) (1)edge[endlabel=$b$]++(-60:0.4) (0)edge[endlabel=$c$]++(120:0.4) (0)edge[endlabel=$d$]++(-120:0.4);
\trianglebot{-90}{(1)}{-};
\trianglebot{-30}{(0)}{-};
\end{tikzpicture}
=
\begin{tikzpicture}
\coords{0/0,0|1/0,0.5};
\draw (0)--(1) (1)edge[endlabel=$a$]++(30:0.4) (0)edge[endlabel=$b$]++(-30:0.4) (1)edge[endlabel=$c$]++(150:0.4) (0)edge[endlabel=$d$]++(-150:0.4);
\trianglebot{0}{(1)}{-};
\trianglebot{-60}{(0)}{-};
\end{tikzpicture}
\end{equation}

Another possibility is to associate the same $4$-index tensor to each edge, and contract the tensors if the edges share a common vertex and face. E.g.:
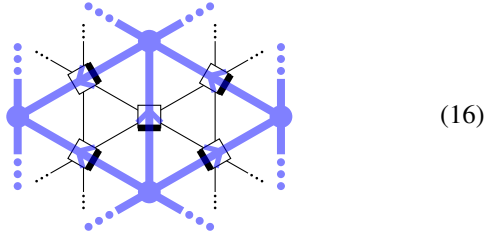
\begin{equation}
\label{eq:tl_example2}
\begin{tikzpicture}
\coords{0/0,0|1/30:2|2/150:2|3/0,2};
\coordinate (e0) at ($(0)!0.5!(1)$);
\coordinate (e1) at ($(0)!0.5!(2)$);
\coordinate (e2) at ($(0)!0.5!(3)$);
\coordinate (e3) at ($(1)!0.5!(3)$);
\coordinate (e4) at ($(2)!0.5!(3)$);
\draw (e0)--(e2) (e1)--(e2) (e0)--(e3) (e1)--(e4) (e2)--(e3) (e2)--(e4) (e0)edge[enddots]++(-30:0.5) (e0)edge[enddots]++(-90:0.5) (e1)edge[enddots]++(-150:0.5) (e1)edge[enddots]++(-90:0.5) (e3)edge[enddots]++(30:0.5) (e3)edge[enddots]++(90:0.5) (e4)edge[enddots]++(150:0.5) (e4)edge[enddots]++(90:0.5);
\squarebot{-60}{(e0)}{+};
\squarebot{60}{(e1)}{+};
\squarebot{0}{(e2)}{+};
\squarebot{60}{(e3)}{+};
\squarebot{120}{(e4)}{+};
\begin{scope}[over]
\draw[overedge,blue] (0)edge[overarr=+](1)(1)edge[overarr=+](3)(3)edge[overarr=+](2)(2)edge[overarr=-](0)(0) (0)edge[overarr=+](3) (0)edge[enddots]++(-30:0.5) (0)edge[enddots]++(-150:0.5) (1)edge[enddots]++(-90:0.5) (1)edge[enddots]++(90:0.5) (2)edge[enddots]++(-90:0.5) (2)edge[enddots]++(90:0.5) (3)edge[enddots]++(30:0.5) (3)edge[enddots]++(150:0.5);
\shapes{\oververtex[blue]}{0,1,2,3};
\end{scope}
\end{tikzpicture}
\end{equation}
Note that due to the symmetry of the situation in the lattice around an edge, the tensors have a index permutation symmetry indicated by the shape. A possible choice of tensor-network move consists of the tensors associated to all edges adjacent to the triangles involved in the Pachner move. E.g. for the Pachner move in Eq.~(\ref{eq:pmove_sample}) we get the following equation:
\begin{equation}
\begin{tikzpicture}
\coords{0/0,0|1/30:2|2/150:2|3/0,2};
\coordinate (e0) at ($(0)!0.5!(1)$);
\coordinate (e1) at ($(0)!0.5!(2)$);
\coordinate (e2) at ($(0)!0.5!(3)$);
\coordinate (e3) at ($(1)!0.5!(3)$);
\coordinate (e4) at ($(2)!0.5!(3)$);
\draw (e0)--(e2) (e1)--(e2) (e0)--(e3) (e1)--(e4) (e2)--(e3) (e2)--(e4) (e0)edge[]++(-30:0.5) (e0)edge[]++(-90:0.5) (e1)edge[]++(-150:0.5) (e1)edge[]++(-90:0.5) (e3)edge[]++(30:0.5) (e3)edge[]++(90:0.5) (e4)edge[]++(150:0.5) (e4)edge[]++(90:0.5);
\squarebot{-60}{(e0)}{+};
\squarebot{60}{(e1)}{+};
\squarebot{0}{(e2)}{+};
\squarebot{60}{(e3)}{+};
\squarebot{120}{(e4)}{+};
\end{tikzpicture}
=
\begin{tikzpicture}[rotate=90]
\coords{0/0,0|1/30:2|2/150:2|3/0,2};
\coordinate (e0) at ($(0)!0.5!(1)$);
\coordinate (e1) at ($(0)!0.5!(2)$);
\coordinate (e2) at ($(0)!0.5!(3)$);
\coordinate (e3) at ($(1)!0.5!(3)$);
\coordinate (e4) at ($(2)!0.5!(3)$);
\draw (e0)--(e2) (e1)--(e2) (e0)--(e3) (e1)--(e4) (e2)--(e3) (e2)--(e4) (e0)edge[]++(-30:0.5) (e0)edge[]++(-90:0.5) (e1)edge[]++(-150:0.5) (e1)edge[]++(-90:0.5) (e3)edge[]++(30:0.5) (e3)edge[]++(90:0.5) (e4)edge[]++(150:0.5) (e4)edge[]++(90:0.5);
\squarebot{-60}{(e0)}{+};
\squarebot{-120}{(e1)}{+};
\squarebot{0}{(e2)}{+};
\squarebot{60}{(e3)}{+};
\squarebot{-60}{(e4)}{+};
\end{tikzpicture}
\end{equation}

Let's consider another TL type with a slightly higher locality length: As in the first example we associate one tensor to each triangle, but now the tensor depends on how many other triangles share common vertices or edges with that triangle. E.g.:
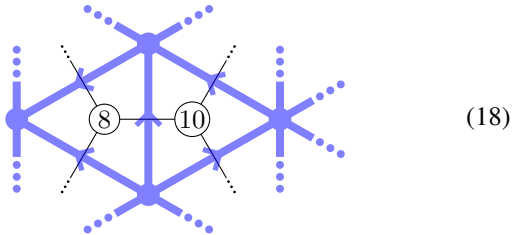
\begin{equation}
\begin{tikzpicture}
\coords{0/0,0|1/30:2|2/150:2|3/0,2};
\coordinate (f0) at ({sqrt(3)/3},1);
\coordinate (f1) at ({-sqrt(3)/3},1);
\draw (f0)--(f1) (f1)edge[enddots]++(120:0.9) (f1)edge[enddots]++(-120:0.9) (f0)edge[enddots]++(60:0.9) (f0)edge[enddots]++(-60:0.9);
\circlenode{$10$}{(f0)};
\circlenode{$8$}{(f1)};
\begin{scope}[over]
\draw[overedge,blue] (0)edge[overarr=+](1)(1)edge[overarr=+](3)(3)edge[overarr=+](2)(2)edge[overarr=-](0)(0) (0)edge[overarr=+](3) (0)edge[enddots]++(-30:0.5) (0)edge[enddots]++(-150:0.5) (1)edge[enddots]++(-90:0.5) (1)edge[enddots]++(90:0.5) (2)edge[enddots]++(-90:0.5) (2)edge[enddots]++(90:0.5) (3)edge[enddots]++(30:0.5) (3)edge[enddots]++(150:0.5) (1)edge[enddots]++(-30:0.5) (1)edge[enddots]++(30:0.5);
\shapes{\oververtex[blue]}{0,1,2,3};
\end{scope}
\end{tikzpicture}
\end{equation}
The tensor-network move for this TL type must contain at least the tensors associated to all triangles that share a common vertex with the triangles involved in the Pachner move. So we get one separate equation for each Pachner move and for each possible depth-two environment of where the move happens. E.g. for a $2$ to $2$ move applied to the lattice patch above we could get an equation like the following:
\begin{equation}
\begin{gathered}
\begin{tikzpicture}[scale=1.5]
\coords{1/-0.3,0|2/0.3,0|3/-0.5,0.5|4/0.5,0.5|5/-0.5,-0.5|6/0.5,-0.5|7/0,0.8|8/0,-0.8|9/-0.9,0|10/1,0.4|11/1.3,0|12/1,-0.4};
\draw (1)--(2) (5)--(8)--(6)--(12)--(11)--(10)--(4)--(7)--(3)--(9)--(5) (3)--(1) (4)--(2) (5)--(1) (6)--(2) (7)--++(90:0.4) (8)--++(-90:0.4) (9)--++(180:0.4) (10)--++(45:0.4) (11)--++(0:0.4) (12)--++(-45:0.4);
\circlenode{$8$}{(1)};
\circlenode{$10$}{(2)};
\circlenode{$8$}{(3)};
\circlenode{$11$}{(4)};
\circlenode{$9$}{(5)};
\circlenode{$12$}{(6)};
\circlenode{$8$}{(7)};
\circlenode{$9$}{(8)};
\circlenode{$9$}{(9)};
\circlenode{$15$}{(10)};
\circlenode{$13$}{(11)};
\circlenode{$12$}{(12)};
\end{tikzpicture}\\
=
\begin{tikzpicture}[scale=1.5]
\coords{1/0,-0.3|2/0,0.3|3/-0.5,0.5|4/0.5,0.5|5/-0.5,-0.5|6/0.5,-0.5|7/0,0.8|8/0,-0.8|9/-0.9,0|10/1,0.4|11/1.3,0|12/1,-0.4};
\draw (1)--(2) (5)--(8)--(6)--(12)--(11)--(10)--(4)--(7)--(3)--(9)--(5) (6)--(1) (3)--(2) (5)--(1) (4)--(2) (7)--++(90:0.4) (8)--++(-90:0.4) (9)--++(180:0.4) (10)--++(45:0.4) (11)--++(0:0.4) (12)--++(-45:0.4);
\circlenode{$10$}{(1)};
\circlenode{$10$}{(2)};
\circlenode{$8$}{(3)};
\circlenode{$11$}{(4)};
\circlenode{$9$}{(5)};
\circlenode{$12$}{(6)};
\circlenode{$7$}{(7)};
\circlenode{$8$}{(8)};
\circlenode{$10$}{(9)};
\circlenode{$16$}{(10)};
\circlenode{$14$}{(11)};
\circlenode{$13$}{(12)};
\end{tikzpicture}
\end{gathered}
\end{equation}

General TL types can have much higher locality length. E.g. we could imagine a TL type similar to the above, just that the tensor at one triangle does not only depend on the number of surrounding triangles, but on the whole patch of lattice distance radius $15$ around the triangle.

\subsubsection*{TL mappings and universality}
A \emph{TL mapping} is a local prescription that transforms TLs of one type into TLs of another type by reshaping the tensor-network and blocking indices. More precisely, in order to define a TL mapping $\mathcal{M}$ from a TL type $\mathcal{A}$ to a TL type $\mathcal{B}$ we need 1) a lattice mapping $\mathcal{M}_L$ from the lattice type of $\mathcal{B}$ to the lattice type of $\mathcal{A}$. Using this mapping we get for every tensor-network geometry of $\mathcal{B}$ a tensor-network geometry of $\mathcal{A}$. Accordingly, we need 2) a \emph{tensor-network mapping} $\mathcal{M}_T$, that is, a prescription that generates a tensor-network geometry of $\mathcal{B}$ by deforming and blocking the associated tensor-network geometry of $\mathcal{A}$.

In order to get the tensor network associated by a lattice $L_{\mathcal{A}}$ by a TL $\mathop{TL}_{\mathcal{A}}$, we apply $\mathcal{M}_L$ to $L_{\mathcal{A}}$, get the associated tensor network by the TL $\mathop{TL}_{\mathcal{B}}$, and deform this tensor network using $\mathcal{M}_T$:
\begin{equation}
\begin{tikzpicture}
\node (la) at (0,0){$L_{\mathcal{A}}$};
\node (lb) at (4,0){$L_{\mathcal{B}}$};
\node (ta) at (0,-1.5){$\mathop{TN}_{\mathcal{A}}$};
\node (tb) at (4,-1.5){$\mathop{TN}_{\mathcal{B}}$};
\draw[->] (la)--node[midway,above]{$\mathcal{M}_L$}(lb);
\draw[<-] (ta)--node[midway,below]{$\mathcal{M}_T$}(tb);
\draw[->] (la)--node[midway,left]{$\mathop{TL}_{\mathcal{A}}$}(ta);
\draw[->] (lb)--node[midway,right]{$\mathop{TL}_{\mathcal{B}}$}(tb);
\end{tikzpicture}
\end{equation}
Thereby the tensor-network moves of $\mathcal{A}$ have to be consistent with the tensor-network moves of $\mathcal{B}$.

Let's consider a few examples for mappings between TL types on $2$tS lattices.

Take for both $\mathcal{A}$ and $\mathcal{B}$ the TL type in Eq.~(\ref{eq:tl_example1}) and for the lattice mapping $\mathcal{M}_L$ the barycentric subdivision in Eq.~(\ref{eq:lattice_mapping_examples}). We can use the following tensor-network mapping: The lattice mapping replaces every triangle with 6 triangles. The tensor-network mapping blocks the $6$ associated tensors into a single one. The indices of the new tensors are composites of two indices of the old tensors:
\begin{equation}
\begin{gathered}
\begin{tikzpicture}
\coords{0/0,0|1/1,0|2/60:1};
\draw[blue] (0)edge[arr=+](1) (0)edge[arr=+](2) (2)edge[arr=+](1);
\shapes{\vertex[blue]}{0,1,2};
\end{tikzpicture}
\quad\rightarrow\quad
\begin{tikzpicture}
\coords{0/-150:1|1/-30:1|2/90:1|m/0,0};
\coordinate (m0) at ($(0)!0.5!(1)$);
\coordinate (m1) at ($(0)!0.5!(2)$);
\coordinate (m2) at ($(2)!0.5!(1)$);
\draw (0)edge[arr=+](m0) (m0)edge[arr=-](1) (0)edge[arr=+](m1) (m1)edge[arr=-](2) (2)edge[arr=+](m2) (m2)edge[arr=-](1) (m)edge[arr=-](0) (m)edge[arr=-](1) (m)edge[arr=-](2) (m)edge[arr=-](m0) (m)edge[arr=-](m1) (m)edge[arr=-](m2);
\shapes{\vertex}{0,1,2,m0,m1,m2,m};
\end{tikzpicture}\\
\begin{tikzpicture}
\coords{0/0,0};
\draw[blue] (0)--++(30:0.7)node[right]{$(b,b')$} (0)--++(150:0.7)node[left]{$(a,a')$} (0)--++(-90:0.7)node[below]{$(c,c')$};
\trianglebot[blue]{0}{(0)}{+};
\end{tikzpicture}=
\begin{tikzpicture}
\coords{0/-180:0.5|1/-120:0.5|2/-60:0.5|3/0:0.5|4/60:0.5|5/120:0.5};
\draw (0)--(1)--(2)--(3)--(4)--(5)--(0) (0)edge[endlabel=$a$]++(150:0.5) (5)edge[endlabel=$a'$]++(150:0.5) (1)edge[endlabel=$c$]++(-90:0.5) (2)edge[endlabel=$c'$]++(-90:0.5) (3)edge[endlabel=$b'$]++(30:0.5) (4)edge[endlabel=$b$]++(30:0.5);
\trianglebot{30}{(0)}{+};
\trianglebot{-150}{(1)}{-};
\trianglebot{150}{(2)}{+};
\trianglebot{-30}{(3)}{-};
\trianglebot{-90}{(4)}{+};
\trianglebot{90}{(5)}{-};
\end{tikzpicture}
\end{gathered}
\end{equation}

As a next example take for $\mathcal{A}$ the TL type in Eq.~(\ref{eq:tl_example1}) and for $\mathcal{B}$ the TL type in Eq.~(\ref{eq:tl_example2}). As lattice mapping $\mathcal{M}_L$ take the \emph{stellar subdivision}, i.e. add one vertex at the center of each triangle and divide the triangle into three triangles by edges connected to the central vertex:
\begin{equation}
\begin{tikzpicture}
\coords{0/0,0|1/30:2|2/150:2|3/0,2};
\coordinate (f0) at ({sqrt(3)/3},1);
\coordinate (f1) at ({-sqrt(3)/3},1);
\draw (0)edge[arr=+](1)(1)edge[arr=+](3)(3)edge[arr=+](2)(2)edge[arr=+](0) (0)edge[arr=+](3) (f0)edge[arr=+](0) (f0)edge[arr=+](3) (f1)edge[arr=+](0) (f1)edge[arr=+](3) (f1)edge[arr=+](2) (f0)edge[arr=+](1);
\draw (0)edge[enddots]++(0:0.3) (0)edge[enddots]++(180:0.3) (0)edge[enddots]++(-90:0.3) (0)edge[enddots]++(-30:0.3) (0)edge[enddots]++(-150:0.3) (1)edge[enddots]++(0:0.3) (1)edge[enddots]++(75:0.3) (1)edge[enddots]++(-75:0.3) (2)edge[enddots]++(-120:0.3) (2)edge[enddots]++(120:0.3) (2)edge[enddots]++(180:0.3) (2)edge[enddots]++(75:0.3) (2)edge[enddots]++(-75:0.3) (3)edge[enddots]++(30:0.3) (3)edge[enddots]++(60:0.3) (3)edge[enddots]++(90:0.3) (3)edge[enddots]++(120:0.3) (3)edge[enddots]++(150:0.3) (3)edge[enddots]++(0:0.3) (3)edge[enddots]++(180:0.3);
\shapes{\vertex}{f0,f1,0,1,2,3};
\begin{scope}[over]
\draw[overedge,blue] (0)edge[overarr=+](1)(1)edge[overarr=+](3)(3)edge[overarr=+](2)(2)edge[overarr=+](0) (0)edge[overarr=+](3) (0)edge[enddots]++(-30:0.5) (0)edge[enddots]++(-150:0.5) (1)edge[enddots]++(0:0.5) (2)edge[enddots]++(-120:0.5) (2)edge[enddots]++(120:0.5) (3)edge[enddots]++(30:0.5) (3)edge[enddots]++(90:0.5) (3)edge[enddots]++(150:0.5);
\shapes{\oververtex[blue]}{0,1,2,3};
\end{scope}
\end{tikzpicture}
\end{equation}
With this lattice mapping we get two triangles of the resulting lattice for each edge of the original lattice. As tensor-network mapping we can block the two associated triangles and use it as the tensor associated to the edge in the original lattice:
\begin{equation}
\begin{gathered}
\begin{tikzpicture}
\coords{0/0,0|1/0,1};
\draw[blue] (0)edge[arr=+](1) (0)edge[enddots]++(30:0.3) (0)edge[enddots]++(150:0.3) (1)edge[enddots]++(-30:0.3) (1)edge[enddots]++(-150:0.3);
\shapes{\vertex[blue]}{0,1};
\end{tikzpicture}
\quad\rightarrow\quad
\begin{tikzpicture}
\coords{0/0,0|1/0,1};
\coordinate (f0) at ({sqrt(3)/6},0.5);
\coordinate (f1) at ({-sqrt(3)/6},0.5);
\draw (0)edge[arr=+](1) (0)edge[arr=-](f0)(f0)edge[arr=+](1) (0)edge[arr=-](f1)(f1)edge[arr=+](1);
\shapes{\vertex}{0,1,f0,f1};
\end{tikzpicture}\\
\begin{tikzpicture}
\coords{0/0,0};
\draw[blue] (0)edge[endlabel=$a$]++(30:0.6) (0)edge[endlabel=$b$]++(-30:0.6) (0)edge[endlabel=$c$]++(150:0.6) (0)edge[endlabel=$d$]++(-150:0.6);
\squarebot[blue]{0}{(0)};
\end{tikzpicture}=
\begin{tikzpicture}
\coords{0/0,0|1/0.5,0};
\draw (0)--(1) (1)edge[endlabel=$a$]++(60:0.4) (1)edge[endlabel=$b$]++(-60:0.4) (0)edge[endlabel=$c$]++(120:0.4) (0)edge[endlabel=$d$]++(-120:0.4);
\trianglebot{150}{(1)}{+};
\trianglebot{-150}{(0)}{-};
\end{tikzpicture}
\end{gathered}
\end{equation}

A TL mapping from a TL type $\mathcal{A}$ to a TL type $\mathcal{B}$ can be seen as a way to encode every TL of type $\mathcal{A}$ into a TL of type $\mathcal{B}$. Such an encoding is faithful when the TL mapping can be inverted by another TL mapping (acting as a left inverse). A (real, $n$tS) TL type $\mathcal{B}$ is \emph{universal} if for every other (real, $n$tS) TL type $\mathcal{A}$ there is a left invertible TL mapping from $\mathcal{A}$ to $\mathcal{B}$. That is, if a TL type is universal, every TL of every other type (with the same lattice type) can be faithfully encoded in a TL of this type.

This paper aims to demonstrate that there are indeed universal (real, $n$tS) TL types, whereas the TL types corresponding to known state-sum constructions are not universal. The main idea to prove that a TL type is universal is to pick a TL mapping with a fine-graining lattice mapping: If the TL type to be mapped has locality length $L$ we can fine-grain by a factor $\gtrsim L$ to obtain a universal TL type with locality length $L \sim 1$. This way one single TL type can effectively describe TL types with arbitrarily high locality length. Note however that the tensors after the mapping arise from blocking large patches of tensor networks, yielding bases for the blocked indices that (in the worst case) grow exponentially in $L$.

In \cite{tensor_lattice} we formulate the concept of a TL for general lattice types (other than $n$tS lattices) and general tensor types (other than real tensors). In general, TL mappings map between different TL types, and universality can be defined with respect to each pair of lattice type and tensor type.

In the following sections we will define universal TL types in $1$ and $2$ dimensions and sketch a universal TL type for arbitrary higher dimensions.

\section{Universality in 1 dimension}
\label{sec:universal_1d}
It is very easy to construct universal TL types in $1$ dimension. Of course it is also possible to write down non-universal TL types, such as one that only associates numbers (i.e. tensors without indices), or no tensors at all. However, already the simplest ``non-degenerate'' constructions that come to mind turn out to be universal TL types. In the following we will describe such a simple universal TL type referred to as 1tSU, where ``U'' stands for ``universal''.

The 1tSU lattices are given by simplicial complexes without edge orientations.

1tSU associates one $2$-index tensor to every edge. The indices of tensors associated to adjacent edges are contracted. E.g.:
\begin{equation}
\begin{tikzpicture}
\coords{0/0,0|1/1,0|2/2,0|3/3,0|m0/0.5,0|m1/1.5,0|m2/2.5,0};
\draw (m0)--(m1)--(m2) (m0)edge[enddots]++(-0.5,0) (m2)edge[enddots]++(0.5,0);
\edgetensor{0}{(m0)};
\edgetensor{0}{(m1)};
\edgetensor{0}{(m2)};
\begin{scope}[over]
\draw[overedge,blue] (0)--(1)--(2)--(3) (0)edge[enddots]++(-0.5,0) (3)edge[enddots]++(0.5,0);
\shapes{\oververtex[blue]}{0,1,2,3};
\end{scope}
\end{tikzpicture}
\end{equation}
The tensor-network move consists of the tensors associated to the edges involved in the move. So we get the following equation for the tensors:
\begin{equation}
\begin{gathered}
\begin{tikzpicture}
\coords{0/0,0|1/1,0|2/2,0};
\draw (0)--(1)--(2) (0)edge[enddots]++(-0.5,0) (2)edge[enddots]++(0.5,0);
\shapes{\vertex}{0,1,2};
\end{tikzpicture}\quad\rightarrow\quad
\begin{tikzpicture}
\coords{0/0,0|1/1,0};
\draw (0)--(1) (0)edge[enddots]++(-0.5,0) (1)edge[enddots]++(0.5,0);
\shapes{\vertex}{0,1};
\end{tikzpicture}\\
\begin{tikzpicture}
\coords{0/0,0|1/1,0};
\draw (0)--(1) (0)--++(-0.5,0) (1)--++(0.5,0);
\edgetensor{0}{(0)};
\edgetensor{0}{(1)};
\end{tikzpicture}=
\begin{tikzpicture}
\coords{0/0,0};
\draw (0)--++(-0.5,0) (0)--++(0.5,0);
\edgetensor{0}{(0)};
\end{tikzpicture}
\end{gathered}
\end{equation}

In order to show the universality of 1tSU we have to find a TL mapping from an arbitrary (1tS, real) TL type $\mathcal{B}$ (with locality length of order $L$) to 1tSU (with locality length of order $1$). We will construct such a mapping now. The lattice mapping from 1tSU to $\mathcal{B}$ is given by fine-graining by a factor of $>2L$, e.g. for $L=2$:
\begin{equation}
\begin{tikzpicture}
\coords{0/0,0|1/2,0|2/4,0};
\foreach \x in {0,0.4,...,3.6}{
\draw (\x,0)--++(0.4,0);
\vertex{(\x,0)};
}
\vertex{(2)};
\draw (0)edge[enddots]++(-0.3,0) (2)edge[enddots]++(0.3,0);
\draw[<->,midlabel=$>2L$] ($(0)+(0,0.4)$)--($(1)+(0,0.4)$);
\begin{scope}[over]
\draw[overedge,blue] (0)--(1)--(2) (0)edge[enddots]++(-0.8,0) (2)edge[enddots]++(0.8,0);
\shapes{\oververtex[blue]}{0,1,2};
\end{scope}
\end{tikzpicture}
\end{equation}
The tensor-network mapping from $\mathcal{B}$ to 1tSU is given by taking all the tensors on the $\mathcal{B}$ lattice corresponding to one unit cell of 1tSU and blocking them into one single tensor:
\begin{equation}
\begin{tikzpicture}
\coords{0/0,0};
\draw (0)--++(-0.5,0) (0)--++(0.5,0);
\edgetensor{0}{(0)};
\end{tikzpicture}
\quad=\quad
\begin{tikzpicture}
\foreach \x in {0,0.4,...,1.6}{
\draw (\x,0)--++(0.4,0);
\vertex{(\x,0)};
}
\vertex{(2,0)};
\end{tikzpicture}
\end{equation}
Note that the right hand side is not tensor-network notation, but symbolizes the tensor-network patch on the shown lattice patch of $\mathcal{B}$.

A Pachner move on the 1tSU lattice corresponds to a sequence of Pachner moves on the corresponding $\mathcal{B}$ lattice. These moves can be chosen to only act within a patch of size $\sim L$ in the center, e.g.:
\begin{equation}
\begin{gathered}
\begin{tikzpicture}
\coords{0/0,0|1/2,0|2/4,0};
\foreach \x in {0,0.4,...,3.6}{
\draw (\x,0)--++(0.4,0);
\vertex{(\x,0)};
}
\vertex{(2)};
\draw (0)edge[enddots]++(-0.3,0) (2)edge[enddots]++(0.3,0);
\begin{scope}[over]
\draw[overedge,blue] (0)--(1)--(2) (0)edge[enddots]++(-0.8,0) (2)edge[enddots]++(0.8,0);
\shapes{\oververtex[blue]}{0,1,2};
\end{scope}
\draw[rounded corners,red,dashed,fill,fill opacity=0.2] (0.7,-0.3)rectangle(3.3,0.3);
\end{tikzpicture}
\\
\rightarrow \quad
\begin{tikzpicture}
\coords{0/0,0|1/2,0};
\foreach \x in {0,0.4,...,1.6}{
\draw (\x,0)--++(0.4,0);
\vertex{(\x,0)};
}
\vertex{(1)};
\draw (0)edge[enddots]++(-0.3,0) (1)edge[enddots]++(0.3,0);
\begin{scope}[over]
\draw[overedge,blue] (0)--(1) (0)edge[enddots]++(-0.8,0) (1)edge[enddots]++(0.8,0);
\shapes{\oververtex[blue]}{0,1};
\end{scope}
\draw[rounded corners,red,dashed,fill,fill opacity=0.2] (0.7,-0.3)rectangle(1.3,0.3);
\end{tikzpicture}
\end{gathered}
\end{equation}
For each Pachner move of $\mathcal{B}$ the corresponding tensor-network move is at most $L$ bigger than the patch where the lattice moves happens. Thus after blocking the following equation holds:
\begin{equation}
\begin{gathered}
\begin{tikzpicture}
\coords{0/0,0|1/2,0|2/4,0};
\foreach \x in {0,0.4,...,3.6}{
\draw (\x,0)--++(0.4,0);
\vertex{(\x,0)};
}
\vertex{(2)};
\begin{scope}[over]
\draw[overedge,blue] (0)--(1)--(2);
\shapes{\oververtex[blue]}{0,1,2};
\end{scope}
\end{tikzpicture}
\\
= \quad
\begin{tikzpicture}
\coords{0/0,0|1/2,0};
\foreach \x in {0,0.4,...,1.6}{
\draw (\x,0)--++(0.4,0);
\vertex{(\x,0)};
}
\vertex{(1)};
\begin{scope}[over]
\draw[overedge,blue] (0)--(1);
\shapes{\oververtex[blue]}{0,1};
\end{scope}
\end{tikzpicture}
\end{gathered}
\end{equation}
This is consistent with the choice of tensor-network moves of 1tSU.

\section{A universal TL type in 2 dimensions}
\label{sec:universal_2d}
In $2$ dimensions it is a bit harder to find universal TL types. One could think that a similar construction as in $1$ dimension could work: Associate one $3$-index tensor to each triangle of a $2$tS lattice and contract indices of adjacent triangles. This does not yield a universal TL type though. The universal TL type that we describe in the following, referred to as $2$tSU (``U'' again stands for ``universal''), turns out to be slightly more complex.

\subsection{Lattices}
\label{sec:2tsu_lattices}
The 2tSU lattices are 2tS lattices with Pachner moves. The edges of the 2tSU lattices are decorated with orientations and dual orientations.

According to Sec.~(\ref{sec:lattices}), 2tS lattices should have a finite set of allowed links for the vertices, i.e. numbers $l$ of faces around a vertex. In principle it suffices to limit $l$ by any large enough number, say $l<100$. In the following we will try to find a set of such numbers $l$ that is as small as possible.

We find that every simplicial complex can be mapped to one with $l=5,6,7$, by the following lattice mapping: We start with a \emph{thickening} of the simplicial complex, i.e. transforming it into a cell complex by replacing every vertex by a $2x$-gon (where $x$ is the number of faces adjacent to the vertex), every edge by a $4$-gon and every face by a $6$-gon. (This is Poincar\'e dual to what is known as \emph{barycentric subdivision}.) E.g.:
\begin{equation}
\begin{tikzpicture}
\coords{0/0,0|1/30:2|2/150:2|3/0,2};
\coordinate (a) at ($(0)+(45:0.8)$);
\coordinate (b) at ($(0)+(105:0.8)$);
\coordinate (c) at ($(1)+(165:0.8)$);
\coordinate (d) at ($(2)+(-15:0.8)$);
\coordinate (e) at ($(3)+(-75:0.8)$);
\coordinate (f) at ($(3)+(-135:0.8)$);
\coordinate (g) at ($(0)+(-15:0.8)$);
\coordinate (h) at ($(0)+(165:0.8)$);
\coordinate (i) at ($(3)+(-15:0.8)$);
\coordinate (j) at ($(3)+(165:0.8)$);
\coordinate (k) at ($(1)+(-135:0.8)$);
\coordinate (l) at ($(1)+(105:0.8)$);
\coordinate (m) at ($(2)+(-75:0.8)$);
\coordinate (n) at ($(2)+(45:0.8)$);
\coordinate (a1) at ($(0)+(75:0.8)$);
\coordinate (b1) at ($(0)+(135:0.8)$);
\coordinate (c1) at ($(1)+(195:0.8)$);
\coordinate (d1) at ($(2)+(15:0.8)$);
\coordinate (e1) at ($(3)+(-45:0.8)$);
\coordinate (f1) at ($(3)+(-105:0.8)$);
\coordinate (g1) at ($(0)+(15:0.8)$);
\coordinate (h1) at ($(0)+(195:0.8)$);
\coordinate (i1) at ($(3)+(15:0.8)$);
\coordinate (j1) at ($(3)+(195:0.8)$);
\coordinate (k1) at ($(1)+(-105:0.8)$);
\coordinate (l1) at ($(1)+(135:0.8)$);
\coordinate (m1) at ($(2)+(-45:0.8)$);
\coordinate (n1) at ($(2)+(75:0.8)$);
\draw (j)--(j1)--(f)--(f1)--(e)--(e1)--(i)--(i1) (k1)--(k)--(c1)--(c)--(l1)--(l) (h1)--(h)--(b1)--(b)--(a1)--(a)--(g1)--(g) (m)--(m1)--(d)--(d1)--(n)--(n1) (a)--(c1) (g1)--(k) (c)--(e1) (l1)--(i) (d1)--(f) (n)--(j1) (m1)--(h) (d)--(b1) (b)--(f1) (a1)--(e);
\draw (h1)edge[enddots]++(-60:0.3) (g)edge[enddots]++(-120:0.3) (k1)edge[enddots]++(0:0.3) (l)edge[enddots]++(0:0.3) (i1)edge[enddots]++(120:0.3) (j)edge[enddots]++(60:0.3) (n1)edge[enddots]++(180:0.3) (m)edge[enddots]++(180:0.3);
\draw (k1)edge[enddots]++(-90:0.3) (g)edge[enddots]++(-30:0.3) (h1)edge[enddots]++(-150:0.3) (m)edge[enddots]++(-90:0.3) (n1)edge[enddots]++(90:0.3) (j)edge[enddots]++(150:0.3) (i1)edge[enddots]++(30:0.3) (l)edge[enddots]++(90:0.3);
\shapes{\vertex}{a,b,c,d,e,f,g,h,i,j,k,l,m,n,a1,b1,c1,d1,e1,f1,g1,h1,i1,j1,k1,l1,m1,n1};
\begin{scope}[over]
\draw[overedge,blue] (0)--(1)--(3)--(2)--(0) (0)--(3) (0)edge[enddots]++(-30:0.5) (0)edge[enddots]++(-150:0.5) (1)edge[enddots]++(-90:0.5) (1)edge[enddots]++(90:0.5) (2)edge[enddots]++(-90:0.5) (2)edge[enddots]++(90:0.5) (3)edge[enddots]++(30:0.5) (3)edge[enddots]++(150:0.5);
\shapes{\oververtex[blue]}{0,1,2,3};
\end{scope}
\end{tikzpicture}
\end{equation}
Then we triangulate each $2x$-gon face (including the $4$-gons and $6$-gons with $x=2,3$) of the thickening in the following way: 1) Choose two vertices opposite to each other in the $2x$-gon. 2) Insert a sequence of $x-1$ edges between the chosen vertices, connected by $x-2$ new vertices. This divides the $2x$-gon into two $2x$-gons. 3) Triangulate the two $2x$-gons in a zigzag manner starting and ending at the chosen vertices. I.e. for a $x=2,3,4$:
\begin{equation}
\begin{gathered}
a)\quad
\begin{tikzpicture}
\coords{0/0,0|1/1,0|2/1,0.7|3/0,0.7};
\draw (0)--(1)--(2)--(3)--(0);
\shapes{\vertex}{0,1,2,3};
\end{tikzpicture}\quad\rightarrow\quad
\begin{tikzpicture}
\coords{0/0,0|1/1,0|2/1,0.7|3/0,0.7};
\draw (0)--(1)--(2)--(3)--(0);
\draw (0)--(2);
\shapes{\vertex}{0,1,2,3};
\end{tikzpicture}\\
b)\quad
\begin{tikzpicture}
\coords{0/0:0.7|1/60:0.7|2/120:0.7|3/180:0.7|4/-120:0.7|5/-60:0.7};
\draw (0)--(1)--(2)--(3)--(4)--(5)--(0);
\shapes{\vertex}{0,1,2,3,4,5};
\end{tikzpicture}\quad\rightarrow\quad
\begin{tikzpicture}
\coords{0/0:0.7|1/60:0.7|2/120:0.7|3/180:0.7|4/-120:0.7|5/-60:0.7|m/0,0};
\draw (0)--(1)--(2)--(3)--(4)--(5)--(0);
\draw (0)--(m) (1)--(m) (2)--(m) (3)--(m) (4)--(m) (5)--(m);
\shapes{\vertex}{0,1,2,3,4,5,m};
\end{tikzpicture}\\
c)\quad
\begin{tikzpicture}
\coords{0/0:0.7|1/45:0.7|2/90:0.7|3/135:0.7|4/180:0.7|5/-135:0.7|6/-90:0.7|7/-45:0.7};
\draw (0)--(1)--(2)--(3)--(4)--(5)--(6)--(7)--(0);
\shapes{\vertex}{0,1,2,3,4,5,6,7};
\end{tikzpicture}\quad\rightarrow\quad
\begin{tikzpicture}
\coords{0/0:0.7|1/45:0.7|2/90:0.7|3/135:0.7|4/180:0.7|5/-135:0.7|6/-90:0.7|7/-45:0.7|m1/0,-0.2|m2/0,0.2};
\draw (0)--(1)--(2)--(3)--(4)--(5)--(6)--(7)--(0);
\draw (3)--(m2)--(4)--(m1)--(5) (1)--(m2)--(0)--(m1)--(7) (2)--(m2)--(m1)--(6);
\shapes{\vertex}{0,1,2,3,4,5,6,7,m1,m2};
\end{tikzpicture}
\end{gathered}
\end{equation}
All of the newly inserted vertices of such a triangulation have $6$ adjacent faces. The vertices at the boundary of the $2x$-gon have $1$ or $2$ adjacent faces for $x=2$, $2$ adjacent faces for $x=3$, and $2$ or $3$ adjacent faces for $x> 3$. Each vertex of the thickening is adjacent to three faces coming from an edge, a triangle and a vertex of the original lattice, i.e. $2x$-gons for $x=2$, $x=3$ and $x\geq 3$ (vertices of the original lattice with less than $3$ adjacent faces can be excluded or easily dealt with). So adding up we get between $l=1+2+2=5$ and $l=2+2+3=7$ adjacent triangles around the vertex. So we have mapped an arbitrary lattice to one with only vertices with $l=5,6,7$.

If we restrict the possible numbers $l$ we also have to restrict the possible Pachner moves, such that the vertices adjacent to the triangles involved in the move have allowed numbers $l$ before and after the move. If we restrict to $l=5,6,7$, the $3$ to $1$ Pachner move is not possible anymore as the vertex that is added/removed in this move has $l=3$. With this move missing our lattice type is not ``topological'' anymore as the $2$ to $2$ Pachner moves cannot change the number of vertices, which is not a topological invariant. In other words, the restricted Pachner moves are not compatible with the lattice mapping above, as there are (non-restricted) Pachner moves that do not correspond to a sequence of restricted Pachner moves after the mapping. Also, it might be necessary to take intermediate steps over lattices with $l>7$ in order to represent (non-restricted) Pachner moves by a sequence of Pachner moves after the mapping.

There are two possible solutions to this problem: 1) We allow different and more complicated moves than Pachner moves. 2) We stick to restricted Pachner moves, but allow a slightly larger range of values for $l$. For us the second approach seems more tractable. As we have seen we need at least $l=3,4,5,6,7$ in order to insert new vertices via a $3$ to $1$ Pachner move. We found sequences of Pachner moves after the mapping for each Pachner move before the mapping that only go over configurations with $l=3,4,5,6,7,8,9,10$, but haven't found a systematic and compact way yet to write them down. It seems plausible that already smaller sets like $l=3,4,5,6,7,8$ are possible.

\subsection{TL type}
The type 2tSU has $3$ kinds of tensors, called \emph{face}, \emph{edge} and \emph{vertex tensor}, which are real tensors with $2$ different index types called \emph{face indices} and \emph{vertex indices}: One associates a face tensor with $3$ face indices to every triangle, one edge tensor with $2$ face and $2$ vertex indices to every edge, and one vertex tensor with $l$ vertex indices to every vertex with $l$-gon link. The face indices of the face tensors are contracted with the face indices of the adjacent edges, and the vertex indices of the vertex tensors are contracted with the vertex indices of the adjacent edges. E.g.:
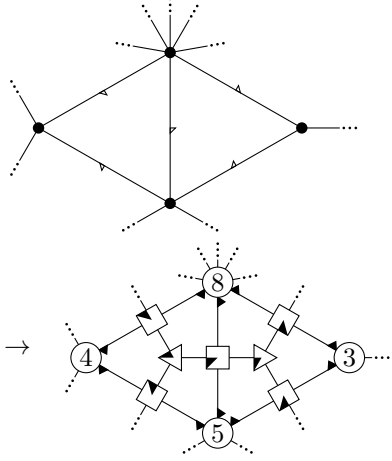
\begin{equation}
\begin{gathered}
\begin{tikzpicture}
\coords{0/0,0|1/30:2|2/150:2|3/0,2};
\draw (0)edge[darr=-+](1) (1)edge[darr=--](3)(3)edge[darr=++](2)(2)edge[darr=--](0) (0)edge[darr=--](3) (0)edge[enddots]++(-30:0.5) (0)edge[enddots]++(-150:0.5) (1)edge[enddots]++(0:0.5) (2)edge[enddots]++(-120:0.5) (2)edge[enddots]++(120:0.5) (3)edge[enddots]++(10:0.5) (3)edge[enddots]++(60:0.5) (3)edge[enddots]++(90:0.5) (3)edge[enddots]++(120:0.5) (3)edge[enddots]++(170:0.5);
\shapes{\vertex}{0,1,2,3};
\end{tikzpicture}\\
\quad\rightarrow\quad
\begin{tikzpicture}
\coords{0/0,0|1/30:2|2/150:2|3/0,2};
\coordinate (e0) at ($(0)!0.5!(1)$);
\coordinate (e1) at ($(0)!0.5!(2)$);
\coordinate (e2) at ($(0)!0.5!(3)$);
\coordinate (e3) at ($(1)!0.5!(3)$);
\coordinate (e4) at ($(2)!0.5!(3)$);
\coordinate (f0) at ({sqrt(3)/3},1);
\coordinate (f1) at ({-sqrt(3)/3},1);
\draw (0)edge[flags=+](e0) (e0)edge[flage=+](1)(1)edge[flags=-](e3)(e3)edge[flage=-](3)(3)edge[flags=+](e4)(e4)edge[flage=+](2)(2)edge[flags=-](e1)(e1)edge[flage=-](0) (0)edge[flags=-](e2)(e2)edge[flage=-](3) (f0)--(e0) (f0)--(e2) (f0)--(e3) (f1)--(e1) (f1)--(e2) (f1)--(e4);
\draw (0)edge[enddots]++(-30:0.3) (0)edge[enddots]++(-150:0.3) (1)edge[enddots]++(0:0.3) (2)edge[enddots]++(-120:0.3) (2)edge[enddots]++(120:0.3) (3)edge[enddots]++(10:0.3) (3)edge[enddots]++(60:0.3) (3)edge[enddots]++(90:0.3) (3)edge[enddots]++(120:0.3) (3)edge[enddots]++(170:0.3) (e0)edge[enddots]++(-60:0.3) (e1)edge[enddots]++(-120:0.3) (e3)edge[enddots]++(60:0.3) (e4)edge[enddots]++(120:0.3);
\squaremark{30}{(e0)}{-};
\squaremark{150}{(e1)}{+};
\squaremark{-90}{(e2)}{+};
\squaremark{-30}{(e3)}{+};
\squaremark{-150}{(e4)}{+};
\circlenode{$5$}{(0)};
\circlenode{$3$}{(1)};
\circlenode{$4$}{(2)};
\circlenode{$8$}{(3)};
\trianglebot{-90}{(f0)}{+};
\trianglebot{-150}{(f1)}{+};
\end{tikzpicture}
\end{gathered}
\end{equation}
The dual edge orientations of the 2tS lattice define edge orientations for the links of its vertices. The vertex tensor depends on those edge orientations of the link. Also the face and edge tensors are sensitive to the (dual) edge orientations. As a consequence they don't have any index permutation symmetries, for which we have included little flags to the shapes. For simplicity, we omit these flags and the (dual) edge orientation below.

The tensor-network moves of 2tSU consist of the tensors associated to all the faces, vertices and edges adjacent to the faces involved in the move. E.g. for the following Pachner move a) we get the axiom b) for the involved tensors:
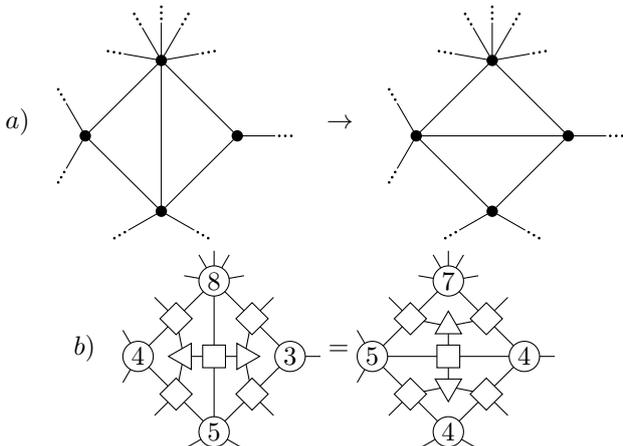
\begin{equation}
\begin{gathered}
a)\quad
\begin{tikzpicture}
\coords{0/0,0|1/1,1|2/-1,1|3/0,2};
\draw (0)--(1)--(3)--(2)--(0) (0)--(3) (0)edge[enddots]++(-30:0.5) (0)edge[enddots]++(-150:0.5) (1)edge[enddots]++(0:0.5) (2)edge[enddots]++(-120:0.5) (2)edge[enddots]++(120:0.5) (3)edge[enddots]++(10:0.5) (3)edge[enddots]++(60:0.5) (3)edge[enddots]++(90:0.5) (3)edge[enddots]++(120:0.5) (3)edge[enddots]++(170:0.5);
\shapes{\vertex}{0,1,2,3};
\end{tikzpicture}
\quad\rightarrow\quad
\begin{tikzpicture}
\coords{0/0,0|1/1,1|2/-1,1|3/0,2};
\draw (0)--(1)--(3)--(2)--(0) (1)--(2) (0)edge[enddots]++(-30:0.5) (0)edge[enddots]++(-150:0.5) (1)edge[enddots]++(0:0.5) (2)edge[enddots]++(-120:0.5) (2)edge[enddots]++(120:0.5) (3)edge[enddots]++(10:0.5) (3)edge[enddots]++(60:0.5) (3)edge[enddots]++(90:0.5) (3)edge[enddots]++(120:0.5) (3)edge[enddots]++(170:0.5);
\shapes{\vertex}{0,1,2,3};
\end{tikzpicture}\\
b)\quad
\begin{tikzpicture}
\coords{0/0,0|1/1,1|2/-1,1|3/0,2};
\coordinate (e0) at ($(0)!0.5!(1)$);
\coordinate (e1) at ($(0)!0.5!(2)$);
\coordinate (e2) at ($(0)!0.5!(3)$);
\coordinate (e3) at ($(1)!0.5!(3)$);
\coordinate (e4) at ($(2)!0.5!(3)$);
\coordinate (f0) at (0.4,1);
\coordinate (f1) at (-0.4,1);
\draw (0)--(e0)--(1)--(e3)--(3)--(e4)--(2)--(e1)--(0) (0)--(e2)--(3) (f0)--(e0) (f0)--(e2) (f0)--(e3) (f1)--(e1) (f1)--(e2) (f1)--(e4);
\draw (0)--++(-30:0.4) (0)--++(-150:0.4) (1)--++(0:0.4) (2)--++(-120:0.4) (2)--++(120:0.4) (3)--++(10:0.4) (3)--++(60:0.4) (3)--++(90:0.4) (3)--++(120:0.4) (3)--++(170:0.4) (e0)--++(-45:0.4) (e1)--++(-135:0.4) (e3)--++(45:0.4) (e4)--++(135:0.4);
\edgetensor{45}{(e0)};
\edgetensor{45}{(e1)};
\edgetensor{0}{(e2)};
\edgetensor{45}{(e3)};
\edgetensor{45}{(e4)};
\circlenode{$5$}{(0)};
\circlenode{$3$}{(1)};
\circlenode{$4$}{(2)};
\circlenode{$8$}{(3)};
\facetensor{-90}{(f0)};
\facetensor{90}{(f1)};
\end{tikzpicture}=
\begin{tikzpicture}
\coords{0/0,0|1/1,1|2/-1,1|3/0,2};
\coordinate (e0) at ($(0)!0.5!(1)$);
\coordinate (e1) at ($(0)!0.5!(2)$);
\coordinate (e2) at ($(1)!0.5!(2)$);
\coordinate (e3) at ($(1)!0.5!(3)$);
\coordinate (e4) at ($(2)!0.5!(3)$);
\coordinate (f0) at (0,1.4);
\coordinate (f1) at (0,0.6);
\draw (0)--(e0)--(1)--(e3)--(3)--(e4)--(2)--(e1)--(0) (1)--(e2)--(2) (f0)--(e4) (f0)--(e2) (f0)--(e3) (f1)--(e1) (f1)--(e0) (f1)--(e2);
\draw (0)--++(-30:0.4) (0)--++(-150:0.4) (1)--++(0:0.4) (2)--++(-120:0.4) (2)--++(120:0.4) (3)--++(10:0.4) (3)--++(60:0.4) (3)--++(90:0.4) (3)--++(120:0.4) (3)--++(170:0.4) (e0)--++(-45:0.4) (e1)--++(-135:0.4) (e3)--++(45:0.4) (e4)--++(135:0.4);
\edgetensor{45}{(e0)};
\edgetensor{45}{(e1)};
\edgetensor{0}{(e2)};
\edgetensor{45}{(e3)};
\edgetensor{45}{(e4)};
\circlenode{$4$}{(0)};
\circlenode{$4$}{(1)};
\circlenode{$5$}{(2)};
\circlenode{$7$}{(3)};
\facetensor{0}{(f0)};
\facetensor{180}{(f1)};
\end{tikzpicture}
\end{gathered}
\end{equation}

In general we get the following equations:
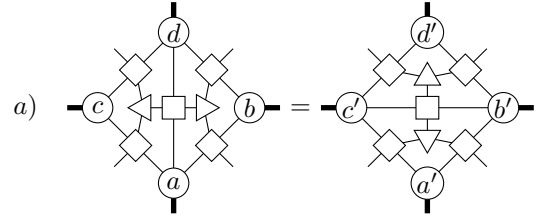
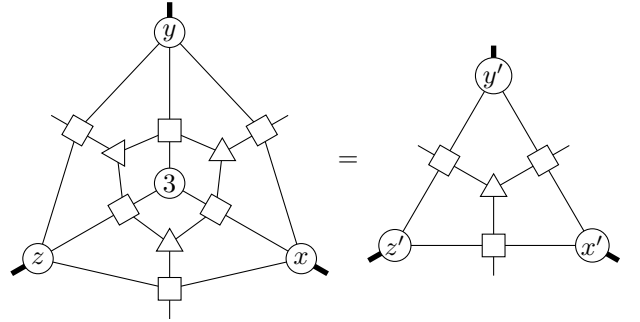
\begin{equation}
\begin{gathered}
a)\quad
\begin{tikzpicture}
\coords{0/0,0|1/1,1|2/-1,1|3/0,2};
\coordinate (e0) at ($(0)!0.5!(1)$);
\coordinate (e1) at ($(0)!0.5!(2)$);
\coordinate (e2) at ($(0)!0.5!(3)$);
\coordinate (e3) at ($(1)!0.5!(3)$);
\coordinate (e4) at ($(2)!0.5!(3)$);
\coordinate (f0) at (0.4,1);
\coordinate (f1) at (-0.4,1);
\draw (0)--(e0)--(1)--(e3)--(3)--(e4)--(2)--(e1)--(0) (0)--(e2)--(3) (f0)--(e0) (f0)--(e2) (f0)--(e3) (f1)--(e1) (f1)--(e2) (f1)--(e4);
\draw[fat] (0)--++(-90:0.4) (1)--++(0:0.4) (2)--++(180:0.4) (3)--++(90:0.4);
\draw (e0)--++(-45:0.4) (e1)--++(-135:0.4) (e3)--++(45:0.4) (e4)--++(135:0.4);
\edgetensor{45}{(e0)};
\edgetensor{45}{(e1)};
\edgetensor{0}{(e2)};
\edgetensor{45}{(e3)};
\edgetensor{45}{(e4)};
\circlenode{$a$}{(0)};
\circlenode{$b$}{(1)};
\circlenode{$c$}{(2)};
\circlenode{$d$}{(3)};
\facetensor{-90}{(f0)};
\facetensor{90}{(f1)};
\end{tikzpicture}=
\begin{tikzpicture}
\coords{0/0,0|1/1,1|2/-1,1|3/0,2};
\coordinate (e0) at ($(0)!0.5!(1)$);
\coordinate (e1) at ($(0)!0.5!(2)$);
\coordinate (e2) at ($(1)!0.5!(2)$);
\coordinate (e3) at ($(1)!0.5!(3)$);
\coordinate (e4) at ($(2)!0.5!(3)$);
\coordinate (f0) at (0,1.4);
\coordinate (f1) at (0,0.6);
\draw (0)--(e0)--(1)--(e3)--(3)--(e4)--(2)--(e1)--(0) (1)--(e2)--(2) (f0)--(e4) (f0)--(e2) (f0)--(e3) (f1)--(e1) (f1)--(e0) (f1)--(e2);
\draw[fat] (0)--++(-90:0.4) (1)--++(0:0.4) (2)--++(180:0.4) (3)--++(90:0.4);
\draw (e0)--++(-45:0.4) (e1)--++(-135:0.4) (e3)--++(45:0.4) (e4)--++(135:0.4);
\edgetensor{45}{(e0)};
\edgetensor{45}{(e1)};
\edgetensor{0}{(e2)};
\edgetensor{45}{(e3)};
\edgetensor{45}{(e4)};
\circlenode{$a'$}{(0)};
\circlenode{$b'$}{(1)};
\circlenode{$c'$}{(2)};
\circlenode{$d'$}{(3)};
\facetensor{0}{(f0)};
\facetensor{180}{(f1)};
\end{tikzpicture}\\
b)\quad
\begin{tikzpicture}
\coords{0/0,0|1/-30:2|2/90:2|3/-150:2|e0/-30:0.7|e1/90:0.7|e2/-150:0.7|e3/30:1.4|e4/150:1.4|e5/-90:1.4|f0/30:0.8|f1/150:0.8|f2/-90:0.8};
\draw (1)--(e3)--(2)--(e4)--(3)--(e5)--(1) (0)--(e0)--(1) (0)--(e1)--(2) (0)--(e2)--(3) (f0)--(e0) (f0)--(e1) (f0)--(e3) (f1)--(e1) (f1)--(e2) (f1)--(e4) (f2)--(e2) (f2)--(e0) (f2)--(e5);
\draw[fat] (1)--++(-30:0.4) (2)--++(90:0.4) (3)--++(-150:0.4);
\draw (e3)--++(30:0.4) (e4)--++(150:0.4) (e5)--++(-90:0.4);
\edgetensor{-30}{(e0)};
\edgetensor{90}{(e1)};
\edgetensor{-150}{(e2)};
\edgetensor{30}{(e3)};
\edgetensor{150}{(e4)};
\edgetensor{-90}{(e5)};
\circlenode{$3$}{(0)};
\circlenode{$x$}{(1)};
\circlenode{$y$}{(2)};
\circlenode{$z$}{(3)};
\facetensor{120}{(f0)};
\facetensor{-30}{(f1)};
\facetensor{0}{(f2)};
\end{tikzpicture}=
\begin{tikzpicture}
\coords{0/0,0|1/-30:1.5|2/90:1.5|3/-150:1.5};
\coordinate (e0) at ($(1)!0.5!(2)$);
\coordinate (e1) at ($(2)!0.5!(3)$);
\coordinate (e2) at ($(3)!0.5!(1)$);
\draw (1)--(e0)--(2)--(e1)--(3)--(e2)--(1) (0)--(e0) (0)--(e1) (0)--(e2);
\draw[fat] (1)--++(-30:0.4) (2)--++(90:0.4) (3)--++(-150:0.4);
\draw (e0)--++(30:0.4) (e1)--++(150:0.4) (e2)--++(-90:0.4);
\edgetensor{30}{(e0)};
\edgetensor{150}{(e1)};
\edgetensor{-90}{(e2)};
\circlenode{$x'$}{(1)};
\circlenode{$y'$}{(2)};
\circlenode{$z'$}{(3)};
\facetensor{0}{(0)};
\end{tikzpicture}
\end{gathered}
\end{equation}
for each set of allowed $l$-values
\begin{equation}
\begin{gathered}
a,b,c,d,a',b',c',d'\\
x,y,z,x',y',z'
\end{gathered}
\end{equation}
with
\begin{equation}
\begin{gathered}
a'=a-1,\quad d'=d-1,\quad b'=b+1,\quad c'=c+1\\
x'=x-1,\quad y'=y-1,\quad z'=z-1
\end{gathered}
\end{equation}
and for each choice of (dual) edge orientations. Here the fat indices are composites of the remaining indices of the vertex tensors.

\subsection{Universality}
We will now show that the TL type 2tSU is indeed universal. In order to do so we need to find a mapping from every (real, 2tS) TL type to 2tSU. Let $\mathcal{B}$ be such a TL type with locality length $L$.

We start by giving the according lattice mapping from 2tSU to $\mathcal{B}$. As in the $1$-dimensional case the key idea is to choose a lattice mapping that fine-grains by a factor of $\gtrsim L$, in order to reduce the locality length from order $L$ to order $1$. To this end we pick a way to fit a block of $\mathcal{B}$ lattice of size $2L$ into a triangle of 2tSU lattice such that the lattice distance between the corners of the triangle is greater than $2L$, e.g. by a regular tiling:
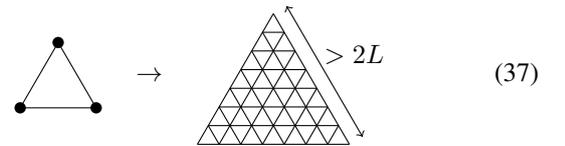
\begin{equation}
\label{eq:mapping_triangle}
\begin{tikzpicture}
\coords{0/0,0|1/1,0|2/60:1};
\draw (0)--(1)--(2)--(0);
\shapes{\vertex}{0,1,2};
\end{tikzpicture}\quad \rightarrow \quad
\begin{tikzpicture}
\coords{0/0,0|1/2,0|2/60:2};
\trianglepattern{(0)}{(1)}{(2)}{7};
\draw[<->,midlabelr=$>2L$,midlabeldistance=0.45cm] ($(1)+(30:0.2)$)--($(2)+(30:0.2)$);
\end{tikzpicture}
\end{equation}

The full lattice mapping consists of the following $3$ steps: 1) Map the 2tSU simplicial complex to a cell complex in the following way: Fatten every vertex with $l$-gon link to a $l$-gon face, and every edge to a $4$-gon face. The vertex-face shares edges with the surrounding edge-faces, but only corners with the surrounding original triangles. E.g.:
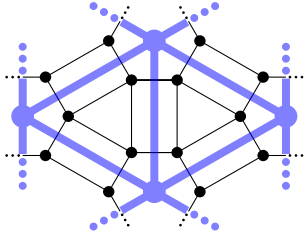
\begin{equation}
\begin{tikzpicture}
\coords{0/0,0|1/30:2|2/150:2|3/0,2};
\coordinate (a) at ($(0)+(60:0.6)$);
\coordinate (b) at ($(0)+(120:0.6)$);
\coordinate (c) at ($(1)+(180:0.6)$);
\coordinate (d) at ($(2)+(0:0.6)$);
\coordinate (e) at ($(3)+(-60:0.6)$);
\coordinate (f) at ($(3)+(-120:0.6)$);
\coordinate (g) at ($(0)+(0:0.6)$);
\coordinate (h) at ($(0)+(180:0.6)$);
\coordinate (i) at ($(3)+(0:0.6)$);
\coordinate (j) at ($(3)+(180:0.6)$);
\coordinate (k) at ($(1)+(-120:0.6)$);
\coordinate (l) at ($(1)+(120:0.6)$);
\coordinate (m) at ($(2)+(-60:0.6)$);
\coordinate (n) at ($(2)+(60:0.6)$);
\draw (h)--(b)--(a)--(g)--(k)--(c)--(l)--(i)--(e)--(f)--(j)--(n)--(d)--(m)--(h) (a)--(c)--(e)--(f)--(d)--(b)--(a) (a)--(e) (b)--(f) (h)edge[enddots]++(-60:0.3) (g)edge[enddots]++(-120:0.3) (k)edge[enddots]++(0:0.3) (l)edge[enddots]++(0:0.3) (i)edge[enddots]++(120:0.3) (j)edge[enddots]++(60:0.3) (n)edge[enddots]++(180:0.3) (m)edge[enddots]++(180:0.3);
\shapes{\vertex}{a,b,c,d,e,f,g,h,i,j,k,l,m,n};
\begin{scope}[over]
\draw[overedge,blue] (0)--(1)--(3)--(2)--(0) (0)--(3) (0)edge[enddots]++(-30:0.5) (0)edge[enddots]++(-150:0.5) (1)edge[enddots]++(-90:0.5) (1)edge[enddots]++(90:0.5) (2)edge[enddots]++(-90:0.5) (2)edge[enddots]++(90:0.5) (3)edge[enddots]++(30:0.5) (3)edge[enddots]++(150:0.5);
\shapes{\oververtex[blue]}{0,1,2,3};
\end{scope}
\end{tikzpicture}
\end{equation}
2) Divide the $4$-gons and $l$-gons coming from the edges and vertices into triangles by adding a vertex to their center. 3) Replace every triangle by the patch of $\mathcal{B}$ lattice as in Eq.~(\ref{eq:mapping_triangle}). E.g.
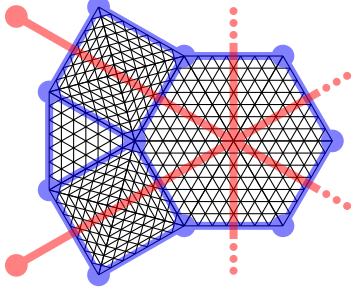
\begin{equation}
\begin{tikzpicture}
\coords{0/0,0|1/0:1.3|2/60:1.3|3/120:1.3|4/180:1.3|5/-120:1.3|6/-60:1.3|x1/150:3.3|x2/-150:3.3};
\coordinate (7) at ($(5)+(-150:1.3)$);
\coordinate (8) at ($(4)+(-150:1.3)$);
\coordinate (9) at ($(4)+(150:1.3)$);
\coordinate (10) at ($(3)+(150:1.3)$);
\coordinate (11) at ($(4)!0.5!(7)$);
\coordinate (12) at ($(3)!0.5!(9)$);
\trianglepattern{(0)}{(1)}{(2)}{7};
\trianglepattern{(0)}{(2)}{(3)}{7};
\trianglepattern{(0)}{(3)}{(4)}{7};
\trianglepattern{(0)}{(4)}{(5)}{7};
\trianglepattern{(0)}{(5)}{(6)}{7};
\trianglepattern{(0)}{(6)}{(1)}{7};
\trianglepattern{(8)}{(9)}{(4)}{7};
\trianglepattern{(11)}{(7)}{(5)}{7};
\trianglepattern{(11)}{(5)}{(4)}{7};
\trianglepattern{(11)}{(4)}{(8)}{7};
\trianglepattern{(11)}{(8)}{(7)}{7};
\trianglepattern{(12)}{(4)}{(3)}{7};
\trianglepattern{(12)}{(3)}{(10)}{7};
\trianglepattern{(12)}{(10)}{(9)}{7};
\trianglepattern{(12)}{(9)}{(4)}{7};
\begin{scope}[over]
\draw[blue,overedge] (1)--(2)--(3)--(4)--(5)--(6)--(1) (4)--(8)--(7)--(5) (3)--(10)--(9)--(4) (8)--(9);
\shapes{\oververtex[blue]}{1,2,3,4,5,6,7,8,9,10};
\end{scope}
\begin{scope}[over]
\draw[overedge,red] (0)--(x1) (0)--(x2) (0)edge[enddots]++(-90:1.3) (0)edge[enddots]++(-30:1.3) (0)edge[enddots]++(30:1.3) (0)edge[enddots]++(90:1.3);
\shapes{\oververtex[red]}{0,x1,x2};
\end{scope}
\end{tikzpicture}
\end{equation}
Here we have draw the original lattice in red, the intermediate fattened lattice in blue, and the final resulting $\mathcal{B}$ lattice in black.

The tensor-network mapping of the TL mapping consists in blocking all the tensors of the $\mathcal{B}$ lattice within the face corresponding to each vertex, edge and triangle of the 2tSU lattice and using it as the corresponding vertex/triangle/face tensor. In order to fix the order of the blocked indices within the composite index we need the orientation and dual orientation of each edge.

Now consider a Pachner move of the 2tSU lattice and the corresponding change of the $\mathcal{B}$ lattice after the mapping, e.g. for a $2\rightarrow 2$ Pachner move:
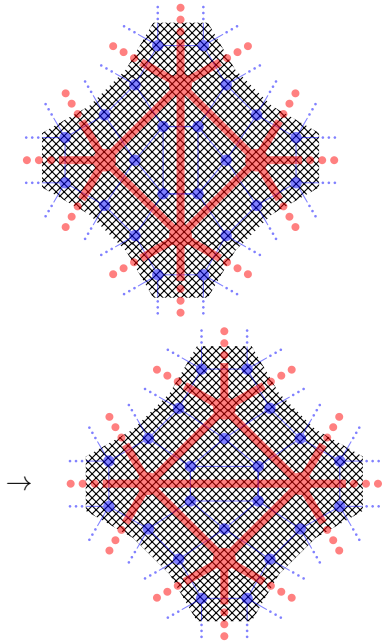
\begin{equation}
\begin{gathered}
\begin{tikzpicture}
\coords{0/0,-1|1/1,0|2/-1,0|3/0,1};
\coordinate (a) at ($(0)+(67.5:0.6)$);
\coordinate (b) at ($(0)+(112.5:0.6)$);
\coordinate (c) at ($(1)+(180:0.4)$);
\coordinate (d) at ($(2)+(0:0.4)$);
\coordinate (e) at ($(3)+(-67.5:0.6)$);
\coordinate (f) at ($(3)+(-112.5:0.6)$);
\coordinate (g) at ($(0)+(0:0.6)$);
\coordinate (h) at ($(0)+(180:0.6)$);
\coordinate (i) at ($(3)+(0:0.6)$);
\coordinate (j) at ($(3)+(180:0.6)$);
\coordinate (k) at ($(1)+(-90:0.6)$);
\coordinate (l) at ($(1)+(90:0.6)$);
\coordinate (m) at ($(2)+(-90:0.6)$);
\coordinate (n) at ($(2)+(90:0.6)$);
\coordinate (o) at ($(1)+(-30:0.6)$);
\coordinate (p) at ($(1)+(30:0.6)$);
\coordinate (q) at ($(2)+(-150:0.6)$);
\coordinate (r) at ($(2)+(150:0.6)$);
\coordinate (s) at ($(0)+(-60:0.6)$);
\coordinate (t) at ($(0)+(-120:0.6)$);
\coordinate (u) at ($(3)+(60:0.6)$);
\coordinate (v) at ($(3)+(120:0.6)$);
\path[hatch] ($1.2*(t)$)--($1.2*(s)$)--($1.2*(g)$)--($1.2*(k)$)--($1.2*(o)$)--($1.2*(p)$)--($1.2*(l)$)--($1.2*(i)$)--($1.2*(u)$)--($1.2*(v)$)--($1.2*(j)$)--($1.2*(n)$)--($1.2*(r)$)--($1.2*(q)$)--($1.2*(m)$)--($1.2*(h)$)--cycle;
\begin{scope}[over]
\draw[blue] (h)--(b)--(a)--(g)--(k)--(c)--(l)--(i)--(e)--(f)--(j)--(n)--(d)--(m)--(h) (a)--(c)--(e)--(f)--(d)--(b)--(a) (a)--(e) (b)--(f) (h)--(t)--(s)--(g) (k)--(o)--(p)--(l) (i)--(u)--(v)--(j) (m)--(q)--(r)--(n) (g)edge[enddots]++(-30:0.3) (k)edge[enddots]++(-60:0.3) (o)edge[enddots]++(0:0.3) (o)edge[enddots]++(-60:0.3) (p)edge[enddots]++(0:0.3) (p)edge[enddots]++(60:0.3) (l)edge[enddots]++(60:0.3) (i)edge[enddots]++(30:0.3) (u)edge[enddots]++(30:0.3) (u)edge[enddots]++(90:0.3) (v)edge[enddots]++(90:0.3) (v)edge[enddots]++(150:0.3) (j)edge[enddots]++(150:0.3) (n)edge[enddots]++(120:0.3) (r)edge[enddots]++(120:0.3) (r)edge[enddots]++(180:0.3) (q)edge[enddots]++(180:0.3) (q)edge[enddots]++(-120:0.3) (m)edge[enddots]++(-120:0.3) (h)edge[enddots]++(-150:0.3) (t)edge[enddots]++(-150:0.3) (t)edge[enddots]++(-90:0.3) (s)edge[enddots]++(-90:0.3) (s)edge[enddots]++(-30:0.3);
\shapes{\vertex[blue]}{a,b,c,d,e,f,g,h,i,j,k,l,m,n,o,p,q,r,s,t,u,v};
\end{scope}
\begin{scope}[over]
\draw[overedge,red] (0)--(1)--(3)--(2)--(0) (0)--(3) (0)edge[enddots]++(-30:0.6) (0)edge[enddots]++(-150:0.6) (1)edge[enddots]++(-60:0.6) (1)edge[enddots]++(60:0.6) (2)edge[enddots]++(-120:0.6) (2)edge[enddots]++(120:0.6) (3)edge[enddots]++(30:0.6) (3)edge[enddots]++(150:0.6) (0)edge[enddots]++(-90:0.6) (1)edge[enddots]++(0:0.6) (2)edge[enddots]++(180:0.6) (3)edge[enddots]++(90:0.6);
\shapes{\oververtex[red]}{0,1,2,3};
\end{scope}
\end{tikzpicture}\\
\quad\rightarrow\quad
\begin{tikzpicture}[rotate=90]
\coords{0/0,-1|1/1,0|2/-1,0|3/0,1};
\coordinate (a) at ($(0)+(67.5:0.6)$);
\coordinate (b) at ($(0)+(112.5:0.6)$);
\coordinate (c) at ($(1)+(180:0.4)$);
\coordinate (d) at ($(2)+(0:0.4)$);
\coordinate (e) at ($(3)+(-67.5:0.6)$);
\coordinate (f) at ($(3)+(-112.5:0.6)$);
\coordinate (g) at ($(0)+(0:0.6)$);
\coordinate (h) at ($(0)+(180:0.6)$);
\coordinate (i) at ($(3)+(0:0.6)$);
\coordinate (j) at ($(3)+(180:0.6)$);
\coordinate (k) at ($(1)+(-90:0.6)$);
\coordinate (l) at ($(1)+(90:0.6)$);
\coordinate (m) at ($(2)+(-90:0.6)$);
\coordinate (n) at ($(2)+(90:0.6)$);
\coordinate (o) at ($(1)+(-30:0.6)$);
\coordinate (p) at ($(1)+(30:0.6)$);
\coordinate (q) at ($(2)+(-150:0.6)$);
\coordinate (r) at ($(2)+(150:0.6)$);
\coordinate (s) at ($(0)+(-60:0.6)$);
\coordinate (t) at ($(0)+(-120:0.6)$);
\coordinate (u) at ($(3)+(60:0.6)$);
\coordinate (v) at ($(3)+(120:0.6)$);
\path[hatch] ($1.2*(t)$)--($1.2*(s)$)--($1.2*(g)$)--($1.2*(k)$)--($1.2*(o)$)--($1.2*(p)$)--($1.2*(l)$)--($1.2*(i)$)--($1.2*(u)$)--($1.2*(v)$)--($1.2*(j)$)--($1.2*(n)$)--($1.2*(r)$)--($1.2*(q)$)--($1.2*(m)$)--($1.2*(h)$)--cycle;
\begin{scope}[over]
\draw[blue] (h)--(b)--(a)--(g)--(k)--(c)--(l)--(i)--(e)--(f)--(j)--(n)--(d)--(m)--(h) (a)--(c)--(e)--(f)--(d)--(b)--(a) (a)--(e) (b)--(f) (h)--(t)--(s)--(g) (k)--(o)--(p)--(l) (i)--(u)--(v)--(j) (m)--(q)--(r)--(n) (g)edge[enddots]++(-30:0.3) (k)edge[enddots]++(-60:0.3) (o)edge[enddots]++(0:0.3) (o)edge[enddots]++(-60:0.3) (p)edge[enddots]++(0:0.3) (p)edge[enddots]++(60:0.3) (l)edge[enddots]++(60:0.3) (i)edge[enddots]++(30:0.3) (u)edge[enddots]++(30:0.3) (u)edge[enddots]++(90:0.3) (v)edge[enddots]++(90:0.3) (v)edge[enddots]++(150:0.3) (j)edge[enddots]++(150:0.3) (n)edge[enddots]++(120:0.3) (r)edge[enddots]++(120:0.3) (r)edge[enddots]++(180:0.3) (q)edge[enddots]++(180:0.3) (q)edge[enddots]++(-120:0.3) (m)edge[enddots]++(-120:0.3) (h)edge[enddots]++(-150:0.3) (t)edge[enddots]++(-150:0.3) (t)edge[enddots]++(-90:0.3) (s)edge[enddots]++(-90:0.3) (s)edge[enddots]++(-30:0.3);
\shapes{\vertex[blue]}{a,b,c,d,e,f,g,h,i,j,k,l,m,n,o,p,q,r,s,t,u,v};
\end{scope}
\begin{scope}[over]
\draw[overedge,red] (0)--(1)--(3)--(2)--(0) (0)--(3) (0)edge[enddots]++(-30:0.6) (0)edge[enddots]++(-150:0.6) (1)edge[enddots]++(-60:0.6) (1)edge[enddots]++(60:0.6) (2)edge[enddots]++(-120:0.6) (2)edge[enddots]++(120:0.6) (3)edge[enddots]++(30:0.6) (3)edge[enddots]++(150:0.6) (0)edge[enddots]++(-90:0.6) (1)edge[enddots]++(0:0.6) (2)edge[enddots]++(180:0.6) (3)edge[enddots]++(90:0.6);
\shapes{\oververtex[red]}{0,1,2,3};
\end{scope}
\end{tikzpicture}
\end{gathered}
\end{equation}
As the mapping above preserves the topology, the change of $\mathcal{B}$ lattice can be achieved by a sequence of (Pachner) moves on the $\mathcal{B}$ lattice. The moves in the sequence only need to act within the patch of $\mathcal{B}$ lattice $P_M$ corresponding to the two triangles involved in the original Pachner move. Now consider the patch of $\mathcal{B}$ lattice $P_T$ corresponding to all the faces of the thickening corresponding to the triangles, edges and vertices involved in the move. We can see from the illustration above that the lattice mapping was constructed in such a way that $P_T$ is bigger than $P_M$ by at least a margin of width $L$. $\mathcal{B}$ has locality length $L$, so if we perform a (Pachner) move within $P_M$, the evaluation of the $\mathcal{B}$ tensor network on $P_T$ remains unchanged. E.g. for the $2$ to $2$ Pachner move above we get the following equation (where the drawn patches symbolize the evaluation of the tensor network on those patches):
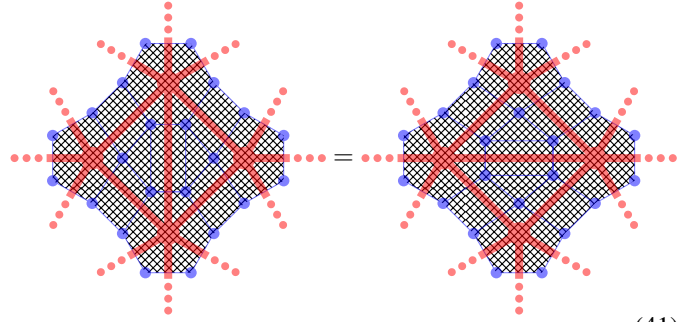
\begin{equation}
\begin{tikzpicture}
\coords{0/0,0|1/1,1|2/-1,1|3/0,2};
\coordinate (a) at ($(0)+(67.5:0.6)$);
\coordinate (b) at ($(0)+(112.5:0.6)$);
\coordinate (c) at ($(1)+(180:0.4)$);
\coordinate (d) at ($(2)+(0:0.4)$);
\coordinate (e) at ($(3)+(-67.5:0.6)$);
\coordinate (f) at ($(3)+(-112.5:0.6)$);
\coordinate (g) at ($(0)+(0:0.6)$);
\coordinate (h) at ($(0)+(180:0.6)$);
\coordinate (i) at ($(3)+(0:0.6)$);
\coordinate (j) at ($(3)+(180:0.6)$);
\coordinate (k) at ($(1)+(-90:0.6)$);
\coordinate (l) at ($(1)+(90:0.6)$);
\coordinate (m) at ($(2)+(-90:0.6)$);
\coordinate (n) at ($(2)+(90:0.6)$);
\coordinate (o) at ($(1)+(-30:0.6)$);
\coordinate (p) at ($(1)+(30:0.6)$);
\coordinate (q) at ($(2)+(-150:0.6)$);
\coordinate (r) at ($(2)+(150:0.6)$);
\coordinate (s) at ($(0)+(-60:0.6)$);
\coordinate (t) at ($(0)+(-120:0.6)$);
\coordinate (u) at ($(3)+(60:0.6)$);
\coordinate (v) at ($(3)+(120:0.6)$);
\fill[hatch] (t)--(s)--(g)--(k)--(o)--(p)--(l)--(i)--(u)--(v)--(j)--(n)--(r)--(q)--(m)--(h)--cycle;
\begin{scope}[over]
\draw[blue] (h)--(b)--(a)--(g)--(k)--(c)--(l)--(i)--(e)--(f)--(j)--(n)--(d)--(m)--(h) (a)--(c)--(e)--(f)--(d)--(b)--(a) (a)--(e) (b)--(f) (h)--(t)--(s)--(g) (k)--(o)--(p)--(l) (i)--(u)--(v)--(j) (m)--(q)--(r)--(n);
\shapes{\vertex[blue]}{a,b,c,d,e,f,g,h,i,j,k,l,m,n,o,p,q,r,s,t,u,v};
\end{scope}
\begin{scope}[over]
\draw[red,overedge] (0)--(1)--(3)--(2)--(0) (0)--(3) (0)edge[enddots]++(-30:0.6) (0)edge[enddots]++(-150:0.6) (1)edge[enddots]++(-60:0.6) (1)edge[enddots]++(60:0.6) (2)edge[enddots]++(-120:0.6) (2)edge[enddots]++(120:0.6) (3)edge[enddots]++(30:0.6) (3)edge[enddots]++(150:0.6) (0)edge[enddots]++(-90:0.6) (1)edge[enddots]++(0:0.6) (2)edge[enddots]++(180:0.6) (3)edge[enddots]++(90:0.6);
\shapes{\oververtex[red]}{0,1,2,3};
\end{scope}
\end{tikzpicture}
=
\begin{tikzpicture}[rotate=90]
\coords{0/0,0|1/1,1|2/-1,1|3/0,2};
\coordinate (a) at ($(0)+(67.5:0.6)$);
\coordinate (b) at ($(0)+(112.5:0.6)$);
\coordinate (c) at ($(1)+(180:0.4)$);
\coordinate (d) at ($(2)+(0:0.4)$);
\coordinate (e) at ($(3)+(-67.5:0.6)$);
\coordinate (f) at ($(3)+(-112.5:0.6)$);
\coordinate (g) at ($(0)+(0:0.6)$);
\coordinate (h) at ($(0)+(180:0.6)$);
\coordinate (i) at ($(3)+(0:0.6)$);
\coordinate (j) at ($(3)+(180:0.6)$);
\coordinate (k) at ($(1)+(-90:0.6)$);
\coordinate (l) at ($(1)+(90:0.6)$);
\coordinate (m) at ($(2)+(-90:0.6)$);
\coordinate (n) at ($(2)+(90:0.6)$);
\coordinate (o) at ($(1)+(-30:0.6)$);
\coordinate (p) at ($(1)+(30:0.6)$);
\coordinate (q) at ($(2)+(-150:0.6)$);
\coordinate (r) at ($(2)+(150:0.6)$);
\coordinate (s) at ($(0)+(-60:0.6)$);
\coordinate (t) at ($(0)+(-120:0.6)$);
\coordinate (u) at ($(3)+(60:0.6)$);
\coordinate (v) at ($(3)+(120:0.6)$);
\fill[hatch] (t)--(s)--(g)--(k)--(o)--(p)--(l)--(i)--(u)--(v)--(j)--(n)--(r)--(q)--(m)--(h)--cycle;
\begin{scope}[over]
\draw[blue] (h)--(b)--(a)--(g)--(k)--(c)--(l)--(i)--(e)--(f)--(j)--(n)--(d)--(m)--(h) (a)--(c)--(e)--(f)--(d)--(b)--(a) (a)--(e) (b)--(f) (h)--(t)--(s)--(g) (k)--(o)--(p)--(l) (i)--(u)--(v)--(j) (m)--(q)--(r)--(n);
\shapes{\vertex[blue]}{a,b,c,d,e,f,g,h,i,j,k,l,m,n,o,p,q,r,s,t,u,v};
\end{scope}
\begin{scope}[over]
\draw[red,overedge] (0)--(1)--(3)--(2)--(0) (0)--(3) (0)edge[enddots]++(-30:0.6) (0)edge[enddots]++(-150:0.6) (1)edge[enddots]++(-60:0.6) (1)edge[enddots]++(60:0.6) (2)edge[enddots]++(-120:0.6) (2)edge[enddots]++(120:0.6) (3)edge[enddots]++(30:0.6) (3)edge[enddots]++(150:0.6) (0)edge[enddots]++(-90:0.6) (1)edge[enddots]++(0:0.6) (2)edge[enddots]++(180:0.6) (3)edge[enddots]++(90:0.6);
\shapes{\oververtex[red]}{0,1,2,3};
\end{scope}
\end{tikzpicture}
\end{equation}
Thus the tensor-network moves of $\mathcal{B}$ are compatible with the tensor-network moves of 2tSU. So we have found a TL mapping from any TL type $\mathcal{B}$ on 2tS lattices (or other types in the same class) to 2tSU.

\section{Higher dimensions}
\label{sec:universal_higherd}
The construction of the universal TL type 2tSU can be generalized to obtain universal TL types $n$tSU on $n$tS lattices, for arbitrary $n$.

$n$tSU associates tensors to every $x$-simplex, for all $0\leq x\leq n$. The associated tensor depends on the link of the $x$-simplex. If a $(x-1)$-simplex is part of a $x$-simplex they have a contracted index between them. The basis of those indices can depend on $x$. The branching structure is not needed. Instead, every $0<x<n$-simplex is decorated with a favorite adjacent $(x+1)$-simplex and a favorite adjacent $(x-1)$-simplex. The tensors associated to a simplex depend on those favorite adjacent simplices for all adjacent simplices.

The tensor-network moves for a Pachner move consist of the tensors associated to all simplices adjacent to the $n$-simplices involved in the Pachner move.

Also the proof of universality, i.e. the construction of the TL mapping from an arbitrary ($n$tS, real) TL type $\mathcal{B}$ to $n$tSU, is analogous to the case of 2tSU TLs in Sec.~(\ref{sec:universal_2d}). Again the lattice mapping consists of three steps: 1) Construct the \emph{semi-thickening} of the $n$tSU simplicial complex, which is a cell complex. It is obtained by replacing every $x$-simplex by a $n$-cell, namely the $x$-simplex times the $(n-x)$-cell whose boundary is the cell complex Poincar\'e dual to the link of the $x$-simplex. Two such $n$-cells share a common $(n-1)$-cell when they correspond to an adjacent pair of $x$-simplex and $(x-1)$-simplex. 2) Replace each $n$-cell by the original $x$-simplex times the stellar cone of the barycentric subdivision of the dual lattice of the link of the $x$-simplex. This divides the $n$-cells of step 1) into smaller $n$-cells. 3) For each possible shape a $n$-cell from step 2) choose a triangulation that is as fine as the locality length $L$ of $\mathcal{B}$. Replace each $n$-cell by the corresponding fine-grained triangulation.

The tensor-network mapping consists in blocking the patches of $\mathcal{B}$ tensor network associated to the $n$-cells from step 1) and using them as the tensor associated to the corresponding simplex. In order to determine the order of the indices within each composite index we can use the favorite adjacent simplices.

When performing a Pachner move on a $n$tSU lattice, the $\mathcal{B}$ lattice changes only inside the patch $P_M$ given by the $n$tSU $n$-simplices involved in the move. So the $n$tSU Pachner move corresponds to a sequence of $\mathcal{B}$ Pachner moves within $P_M$. Consider the patch $P_T$ consisting $n$-cells from step 1) associated to all $x$-simplices adjacent to the $n$-simplices involved in the move. $P_T$ is larger than $P_M$ by at least a margin of width $L$. As $\mathcal{B}$ has locality length $L$, the sequence of $\mathcal{B}$ Pachner moves doesn't change the evaluation of the tensor network on $P_T$. So the tensor-network moves of $n$tSU and $\mathcal{B}$ are consistent under the TL mapping.

\section{Topological boundaries and chiral phases}
\label{sec:top_boundary}
Non-chiral topological phases of matter in $2+1$ dimensions have been classified in a direct physical way via Levin-Wen models \cite{Levin2004}. These take unitary fusion categories as input and yield exactly solvable microscopic models representing these phases. They are nothing but a Hamiltonian formulation of the much older Turaev-Viro state-sum construction \cite{Turaev1992}. A generalization of those models to chiral phases is still lacking, one major obstruction being the absence of commuting-projector Hamiltonians for chiral phases \cite{Kapustin2018}.

Another very common approach is the classification of topological phases via their ``anyon statistics'' given by modular fusion categories. This is not a classification on a direct physical level, however. Whereas it seems to be the case that different anyon statistics are in fact in one-to-one correspondence with different microscopic physical phases in $2+1$ dimensions, this is not proven. (In the language of axiomatic TQFTs, this roughly corresponds to the question whether every $3-2-1$-extended TQFT can be extended down to points in a unique way.) Also it is unclear how ``anyon statistics'' can be generalized to higher dimensions, and whether the same assumption will hold there.

A direct physical classification of general (including chiral) topological phases is one of the most important open problems in the classification of phases of matter. In this section we argue how universal TL types could be the key to solving this problem. In Sec.~(\ref{sec:boundary},\ref{sec:hamiltonian},\ref{sec:tensor_network_gs}) we show that the existence of a gapped boundary, a commuting-projector Hamiltonian, and a tensor-network representation of ground states, are direct implications of the simple form of known state-sum constructions, which do not need to hold for universal TL types. In Sec.~(\ref{sec:chiral}) we will see how this is compatible with chiral topological phases.
\subsection{State-sum constructions with boundary}
\label{sec:boundary}
In this section we will define state-sums on topological $n$-manifolds with boundary. We will show that non-universal ($n$tS, real) TL types similar to the ones known from the literature can always be extended to $n$-dimensional simplicial complexes with boundary.

Intuitively, a \emph{$n$-dimensional simplicial complex with boundary} (short \emph{$n$tbS lattice}) is a decomposition of a $n$-manifold with boundary into $n$-simplices. The $(n-1)$-simplices in the interior connect two $n$-simplices. All $(n-1)$-simplices at the boundary are only connected to one $n$-simplex in the interior and form themselves a $(n-1)$tS lattice. In the following we will draw the simplices in the bulk in blue and those of the boundary in black, e.g.:
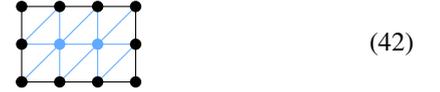
\begin{equation}
\begin{tikzpicture}
\coords{0/0,0|1/0.5,0|2/1,0|3/1.5,0|4/0,0.5|5/0.5,0.5|6/1,0.5|7/1.5,0.5|8/0,1|9/0.5,1|10/1,1|11/1.5,1};
\draw[ctlc1] (4)--(5)--(6)--(7) (1)--(5)--(9) (2)--(6)--(10) (0)--(5) (1)--(6) (2)--(7) (4)--(9) (5)--(10) (6)--(11);
\draw[ctlc0] (0)--(1)--(2)--(3)--(7)--(11)--(10)--(9)--(8)--(4)--(0);
\shapes{\vertex[ctlc0]}{0,1,2,3,7,11,10,9,8,4,0};
\shapes{\vertex[ctlc1]}{5,6};
\end{tikzpicture}
\end{equation}
where we suppressed the edge orientations.

A \emph{boundary Pachner move} changes a $n$tbS lattice locally near the boundary in the following way: It performs a Pachner move on the boundary $(n-1)$tS lattice by attaching or removing a $n$-simplex to the boundary. E.g.:
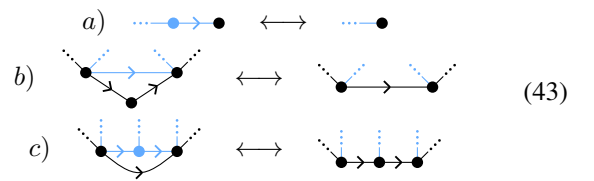
\begin{equation}
\begin{gathered}
a)\quad
\begin{tikzpicture}
\coords{0/0,0|1/0.6,0};
\draw[ctlc1,arr=+] (0)--(1);
\draw[ctlc1] (0)edge[enddots]++(-0.3,0);
\vertex[ctlc0]{(1)};
\vertex[ctlc1]{(0)};
\end{tikzpicture}
\quad \longleftrightarrow\quad
\begin{tikzpicture}
\coords{0/0,0};
\draw[ctlc1] (0)edge[enddots]++(-0.3,0);
\vertex[ctlc0]{(0)};
\end{tikzpicture}\\
b)\quad
\begin{tikzpicture}
\coords{0/0,0|1/0.6,-0.4|2/1.2,0};
\draw[ctlc1] (0)edge[arr=+](2);
\draw[ctlc0] (0)edge[arr=+](1) (1)edge[arr=+](2) (0)edge[enddots]++(135:0.2) (2)edge[enddots]++(45:0.2);
\draw[ctlc1] (0)edge[enddots]++(45:0.2) (2)edge[enddots]++(135:0.2);
\shapes{\vertex[ctlc0]}{0,1,2};
\end{tikzpicture}
\quad \longleftrightarrow \quad
\begin{tikzpicture}
\coords{0/0,0|2/1.2,0};
\draw[ctlc0] (0)edge[arr=+](2);
\draw[ctlc0] (0)edge[enddots]++(135:0.2) (2)edge[enddots]++(45:0.2);
\draw[ctlc1] (0)edge[enddots]++(45:0.2) (2)edge[enddots]++(135:0.2);
\shapes{\vertex[ctlc0]}{0,2};
\end{tikzpicture}\\
c)\quad
\begin{tikzpicture}
\coords{0/0,0|1/0.5,0|2/1,0};
\draw (0)edge[arr=+,ctlc1](1) (1)edge[arr=+,ctlc1](2) (0)edge[arr=+,bend right=40,looseness=1.5,ctlc0](2);
\draw[ctlc0] (0)edge[enddots]++(135:0.2) (2)edge[enddots]++(45:0.2);
\draw[ctlc1] (0)edge[enddots]++(90:0.2) (2)edge[enddots]++(90:0.2) (1)edge[enddots]++(90:0.2);
\shapes{\vertex[ctlc0]}{0,2};
\vertex[ctlc1]{(1)};
\end{tikzpicture}
\quad \longleftrightarrow \quad
\begin{tikzpicture}
\coords{0/0,0|1/0.5,0|2/1,0};
\draw (0)edge[arr=+,ctlc0](1) (1)edge[arr=+,ctlc0](2);
\draw[ctlc0] (0)edge[enddots]++(135:0.2) (2)edge[enddots]++(45:0.2);
\draw[ctlc1] (0)edge[enddots]++(90:0.2) (2)edge[enddots]++(90:0.2) (1)edge[enddots]++(90:0.2);
\shapes{\vertex[ctlc0]}{0,1,2};
\end{tikzpicture}
\end{gathered}
\end{equation}
a) shows a boundary Pachner move in $1$ dimension, whereas b) and c) show boundary Pachner moves in $2$ dimensions.

$n$tbS lattices represent a lattice type of a different class than $n$tS lattices. There is a lattice mapping from $n$tS lattices to $n$tbS lattices: Every simplicial complex is also a simplicial complex with (empty) boundary. However there is no obvious lattice mapping from $n$tbS lattices to $n$tS lattices that would act as a left inverse of the former mapping, as there is no way to get rid of the boundary.

As discussed in the end of Sec.~(\ref{sec:lattices}), simplicial complexes can be modeled by two different lattice types: $n$tS where we control the links and $n$taS where links can be arbitrarily large. For $n$taS, there is the following mapping from $n$tbS to $n$taS: Add to each boundary $(n-1)$-simplex another adjacent $n$-simplex. The $n$-simplices associated to adjacent boundary $(n-1)$-simplices are adjacent to each other. In other words we close the simplicial complex by gluing a stellar cone to each boundary component. E.g. consider the mapping for some lattices in $2$ dimensions:
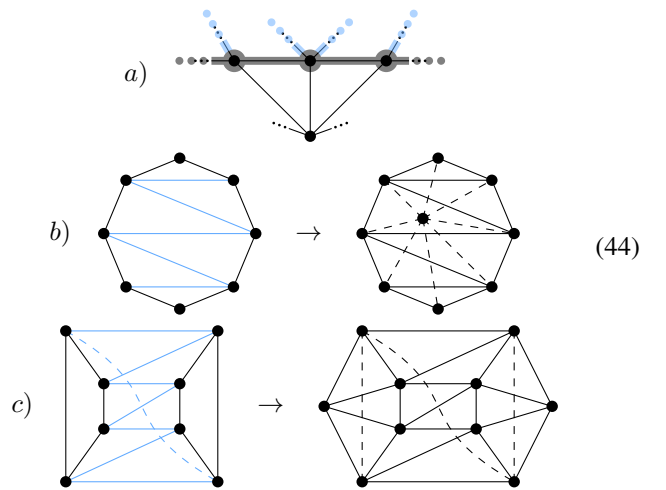
\begin{equation}
\label{eq:boundary_mapping}
\begin{gathered}
a)\quad
\begin{tikzpicture}
\coords{0/0,0|1/1,0|2/2,0|m/1,-1};
\draw[black] (0)--(1)--(2) (0)--(m) (1)--(m) (2)--(m) (0)edge[enddots]++(120:0.3) (1)edge[enddots]++(135:0.3) (1)edge[enddots]++(45:0.3) (2)edge[enddots]++(60:0.3) (0)edge[enddots]++(180:0.3) (2)edge[enddots]++(0:0.3) (m)edge[enddots]++(20:0.3) (m)edge[enddots]++(160:0.3);
\shapes{\vertex[black]}{0,1,2,m};
\begin{scope}[over]
\draw[overedge,ctlc0] (0)--(1)--(2) (0)edge[enddots]++(180:0.3) (2)edge[enddots]++(0:0.3);
\draw[overedge,ctlc1] (0)edge[enddots]++(120:0.3) (1)edge[enddots]++(135:0.3) (1)edge[enddots]++(45:0.3) (2)edge[enddots]++(60:0.3);
\shapes{\oververtex[ctlc0]}{0,1,2};
\end{scope}
\end{tikzpicture}\\
b)\quad
\begin{tikzpicture}
\coords{0/45:1|1/90:1|2/135:1|3/180:1|4/-135:1|5/-90:1|6/-45:1|7/0:1};
\draw[ctlc0] (0)--(1)--(2)--(3)--(4)--(5)--(6)--(7)--(0);
\draw[ctlc1] (0)--(2)--(7)--(3)--(6)--(4);
\shapes{\vertex[ctlc0]}{0,1,2,3,4,5,6,7};
\end{tikzpicture}
\quad\rightarrow\quad
\begin{tikzpicture}
\coords{0/45:1|1/90:1|2/135:1|3/180:1|4/-135:1|5/-90:1|6/-45:1|7/0:1|m/-0.2,0.2};
\draw (0)--(1)--(2)--(3)--(4)--(5)--(6)--(7)--(0);
\draw (0)--(2)--(7)--(3)--(6)--(4);
\draw[backedge] (m)--(0) (m)--(1) (m)--(2) (m)--(3) (m)--(4) (m)--(5) (m)--(6) (m)--(7);
\shapes{\vertex}{0,1,2,3,4,5,6,7,m};
\end{tikzpicture}\\
c)\quad
\begin{tikzpicture}
\coords{0/0,0|1/2,0|2/0,2|3/2,2|4/0.5,0.7|5/1.5,0.7|6/0.5,1.3|7/1.5,1.3};
\draw[ctlc0] (0)--(4)--(6)--(2)--(0) (1)--(3)--(7)--(5)--(1);
\draw[ctlc1] (0)--(1) (2)--(3) (4)--(5) (6)--(7) (0)--(5) (4)--(7) (6)--(3);
\draw[ctlc1,backedge] (2)to[out=-30,in=150,looseness=1.5](1);
\shapes{\vertex[ctlc0]}{0,1,2,3,4,5,6,7};
\end{tikzpicture}
\quad\rightarrow\quad
\begin{tikzpicture}
\coords{0/0,0|1/2,0|2/0,2|3/2,2|4/0.5,0.7|5/1.5,0.7|6/0.5,1.3|7/1.5,1.3|m1/-0.5,1|m2/2.5,1};
\draw[ctlc0] (0)--(4)--(6)--(2) (3)--(7)--(5)--(1);
\draw[ctlc0] (0)--(1) (2)--(3) (4)--(5) (6)--(7) (0)--(5) (4)--(7) (6)--(3) (m1)--(0) (m1)--(2) (m1)--(4) (m1)--(6) (m2)--(1) (m2)--(3) (m2)--(5) (m2)--(7);
\draw[ctlc0,backedge] (2)to[out=-30,in=150,looseness=1.5](1) (1)--(3) (0)--(2);
\shapes{\vertex[ctlc0]}{0,1,2,3,4,5,6,7,m1,m2};
\end{tikzpicture}
\end{gathered}
\end{equation}
a) shows the lattice mapping schematically on two boundary edges of a 2tbS lattice. Here we changed the color scheme: The semi-transparent lattice in blue and black corresponds to the original 2tbS lattice whereas the black lattice is the 2taS lattice resulting from the mapping. b) shows a 2tbS lattice with disk topology yielding a 2taS lattice with sphere topology. c) shows a 2tbS lattice with annulus topology yielding a 2taS lattice with sphere topology.

This construction is not possible with $n$tS instead of $n$taS, as the link of the central vertex of the stellar cone is the $(n-1)$-dimensional simplicial complex forming the boundary, and thus can be arbitrarily large.

Now consider, for arbitrary dimensions $n$, the following TL type \emph{$n$tSX} on $n$tS: Associate one tensor to each $n$-simplex with one index for each of its $(n-1)$-simplices. Indices of neighboring $n$-simplices for the same $(n-1)$-simplex are contracted. The tensor-network moves consist of the tensors associated to the $n$-simplices involved in the Pachner moves. Eq.~(\ref{eq:tl_example1}) shows this TL type in $2$ dimensions. Up to technical details, this is equivalent to the state-sum construction in \cite{Fukuma1992} for $n=2$ and the Turaev-Viro state-sum \cite{Turaev1992} for $n=3$.

The boundary analogue of $n$tSX is the TL type \emph{$n$tbSX} on $n$tbS: In the interior we associate tensors to $n$-simplices, just as for $n$tSX. Additionally, we associate one tensor to each boundary $(n-1)$-simplex. Each of those tensors shares one contracted index with the tensors associated to the adjacent boundary $(n-1)$-simplices, and one with the adjacent bulk $n$-simplex. The tensor-network moves for boundary Pachner moves consist of the tensors associated to the $n$-simplices and boundary $(n-1)$-simplices directly involved in the move. Up to technical details this is equivalent to the state-sum construction for topological manifolds with boundary in \cite{Lauda2006} in $2$ dimensions, and the models for gapped boundaries in \cite{Kitaev2011} in $3$ dimensions.

The TL type $n$tSX can be defined on $n$taS, as there are no tensors or contractions associated to vertices or other $x$ simplices with $x<n-1$. We can extend the lattice mapping from $n$tbS to $n$taS to a TL mapping from $n$tSX to $n$tbSX. The tensor-network mapping consists of taking the tensor associated to the $n$-simplex that was added to a boundary $(n-1)$-simplex as the tensor associated to this boundary $(n-1)$-simplex.

So we found that for every $n$tSX TL there is a $n$tbSX TL that contains the former TL by restricting to $n$-manifolds without boundary. In other words, every $n$tSX TL has a standard topological boundary.

Note that the construction above depends on the specific structure of $n$tSX TLs and is not generalizable to arbitrary TL types on $n$tS, if they cannot be defined on $n$taS. Consider for example the universal TL type 2tSU: It cannot be defined on 2taS, as it has tensors associated to the vertices. We are only given these vertex tensors for a finite number of links numbers $l$ (see Sec.~(\ref{sec:2tsu_lattices})), and the tensor-network is undefined at vertices with arbitrary link numbers, as those arising as center of the stellar cone in the mapping in Eq.~(\ref{eq:boundary_mapping}).

In fact one can define infinite series of vertex tensors for arbitrarily large links (though a fixed TL type must per construction restrict to a finite number of tensors). For TLs that can be put onto $n$taS lattices (and thus have a topological boundary), these tensors can be written as a $(n-1)$-dimensional tensor-network with open indices (known as \emph{projected entangled-pair state}, short \emph{PEPS}, with constant \emph{bond dimension}). For TLs that do not have a topological boundary, the sequence of vertex tensors cannot be written as MPSs with a constant bond dimension. Instead this bond dimension grows with the size of the links (and therefore with the number of indices of the vertex tensors).

\subsection{Commuting-projector Hamiltonians}
\label{sec:hamiltonian}
In the context of quantum mechanics, the lattices of a TL represent an euclidean space-time, and the tensor-networks represent the imaginary time evolution of a quantum system in such a space-time. Physical spaces are given by co-dimension $1$ cuts through a space-time, and the Hilbert space for such a cut is given by the vector space formed by the cut-open indices.

For $n$tSX TLs the physical spaces are given by $(n-1)$tS lattices, and the corresponding Hilbert space has one degree of freedom for each $(n-1)$-simplex. This degree of freedom is a qu-$B$-it with Hilbert space $\mathbb{C}^B$, where $B$ is the basis of the indices of the $n$tSX TL. A commuting-projector Hamiltonian for these quantum models can be constructed as follows:

Consider a vertex $v$ of a physical space $S$. Take a patch $X_v$ of $n$tS lattice consisting of one $n$-simplex for each $(n-1)$-simplex adjacent to the vertex, such that two $n$-simplices share a common $(n-1)$-simplex if the two $(n-1)$-simplices share a $(n-2)$-simplex. E.g. in $2$ and $3$ dimensions:
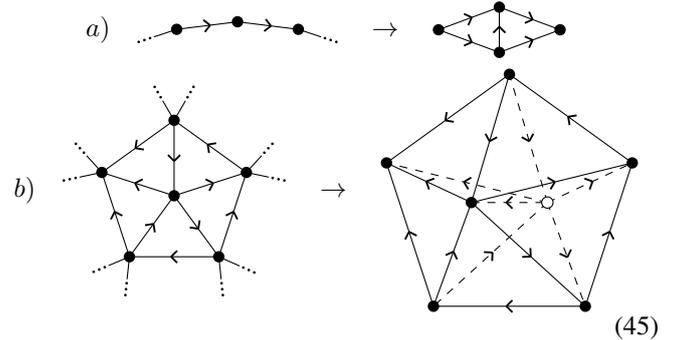
\begin{equation}
\begin{gathered}
a)\quad
\begin{tikzpicture}
\coords{0/0,0|1/0.8,0.1|2/1.6,0};
\draw (0)edge[arr=+](1) (1)edge[arr=+](2) (0)edge[enddots]++(-0.3,-0.1) (2)edge[enddots]++(0.3,-0.1);
\shapes{\vertex}{0,1,2};
\end{tikzpicture}\quad\rightarrow\quad
\begin{tikzpicture}
\coords{3/0,0.4|4/0.8,0.1|5/1.6,0.4|6/0.8,0.7};
\draw (3)edge[arr=+](4) (4)edge[arr=+](5) (3)edge[arr=+](6) (6)edge[arr=+](5) (4)edge[arr=+](6);
\shapes{\vertex}{3,4,5,6};
\end{tikzpicture}\\
b)\quad
\begin{tikzpicture}[plabel]
\coords{0/0,0|1/18:1|2/90:1|3/162:1|4/234:1|5/306:1};
\draw (0)edge[circarr=+](1) (0)edge[circarr=-](2) (0)edge[circarr=+](3) (0)edge[circarr=-](4) (0)edge[circarr=+](5);
\draw (1)edge[circarr=+](2) (2)edge[circarr=+](3) (3)edge[circarr=-](4) (4)edge[circarr=-](5) (5)edge[circarr=+](1);
\draw (1)edge[enddots]++(-12:0.3) (1)edge[enddots]++(48:0.3) (2)edge[enddots]++(60:0.3) (2)edge[enddots]++(120:0.3) (3)edge[enddots]++(192:0.3) (3)edge[enddots]++(132:0.3) (4)edge[enddots]++(264:0.3) (4)edge[enddots]++(203:0.3) (5)edge[enddots]++(276:0.3) (5)edge[enddots]++(336:0.3);
\shapes{\vertex}{0,1,2,3,4,5};
\end{tikzpicture}\quad\rightarrow\quad
\begin{tikzpicture}[plabel]
\coords{0/-0.5,0|1/18:1.7|2/90:1.7|3/162:1.7|4/234:1.7|5/306:1.7|x/0.5,0};
\draw (0)edge[circarr=+](1) (0)edge[circarr=-](2) (0)edge[circarr=+](3) (0)edge[circarr=-](4) (0)edge[circarr=+](5);
\draw[backedge] (x)edge[circarr=+](1) (x)edge[circarr=-](2) (x)edge[circarr=+](3) (x)edge[circarr=-](4) (x)edge[circarr=+](5) (0)edge[circarr=-](x);
\draw (1)edge[circarr=+](2) (2)edge[circarr=+](3) (3)edge[circarr=-](4) (4)edge[circarr=-](5) (5)edge[circarr=+](1);
\shapes{\vertex}{0,1,2,3,4,5};
\backvertex{(x)};
\end{tikzpicture}
\end{gathered}
\end{equation}
a) shows a vertex in a 1tS lattice adjacent to two edges, yielding a patch of 2tS lattice consisting of two triangles. b) shows a vertex in a 2tS lattice adjacent to $5$ triangles, yielding a patch of 3tS lattice consisting of $5$ tetrahedra.

Take the tensor-network patch associated to $X_v$, evaluate it, and interpret the result as a linear map $P_v$ from the indices on the lower half to the indices on the upper half. E.g. in $2$-dimensions we have:
\begin{equation}
\begin{tikzpicture}
\coords{3/0,0.4|4/0.8,0.1|5/1.6,0.4|6/0.8,0.7};
\draw (3)edge[arr=+](4) (4)edge[arr=+](5) (3)edge[arr=+](6) (6)edge[arr=+](5) (4)edge[arr=+](6);
\shapes{\vertex}{3,4,5,6};
\end{tikzpicture}\quad\rightarrow\quad
\begin{tikzpicture}
\coords{0/0,0|1/0.5,0};
\draw (0)--(1) (1)edge[endlabel=$a$]++(60:0.4) (1)edge[endlabel=$b$]++(-60:0.4) (0)edge[endlabel=$c$]++(120:0.4) (0)edge[endlabel=$d$]++(-120:0.4);
\trianglebot{30}{(1)}{+};
\trianglebot{-150}{(0)}{-};
\end{tikzpicture}=
(P_v)_{bd}^{ac}
\end{equation}
Consider the patch $X_v'$ obtained by gluing the lower boundary one copy of $X_v$ to the upper boundary of another copy of $X_v$. Using Pachner moves $X_v'$ can be transformed into $X_v$. E.g.:
\begin{equation}
\label{eq:local_projector}
\begin{gathered}
\begin{tikzpicture}
\coords{3/0,0.4|4/0.8,0.1|5/1.6,0.4|6/0.8,0.7};
\draw (3)edge[arr=+](4) (4)edge[arr=+](5) (3)edge[arr=+](6) (6)edge[arr=+](5) (4)edge[arr=+](6);
\shapes{\vertex}{3,4,5,6};
\end{tikzpicture}\quad\longleftrightarrow\quad
\begin{tikzpicture}
\coords{3/0,0.5|4/0.8,0|5/1.6,0.5|6/0.8,1|7/0.8,0.5};
\draw (3)edge[arr=+](4) (4)edge[arr=+](5) (3)edge[arr=+](6) (6)edge[arr=+](5) (4)edge[arr=+](7) (7)edge[arr=+](6) (3)edge[arr=+](7) (7)edge[arr=+](5);
\shapes{\vertex}{3,4,5,6,7};
\end{tikzpicture}\\
\Rightarrow
P_v=P_v^2
\end{gathered}
\end{equation}
Also using Pachner moves, the lower and upper boundary part of $X_v$ can be swapped:
\begin{equation}
\begin{gathered}
\begin{tikzpicture}
\coords{3/0,0.4|4/0.8,0.1|5/1.6,0.4|6/0.8,0.7};
\draw (3)edge[arr=+](4) (4)edge[arr=+](5) (3)edge[arr=+](6) (6)edge[arr=+](5) (4)edge[arr=+](6);
\shapes{\vertex}{3,4,5,6};
\end{tikzpicture}\quad\longleftrightarrow\quad
\begin{tikzpicture}
\coords{3/0,0.4|4/0.8,0.1|5/1.6,0.4|6/0.8,0.7};
\draw (3)edge[arr=+](4) (4)edge[arr=+](5) (3)edge[arr=+](6) (6)edge[arr=+](5) (4)edge[arr=-](6);
\shapes{\vertex}{3,4,5,6};
\end{tikzpicture}\\
\Rightarrow
P_v=P_v^T
\end{gathered}
\end{equation}
Furthermore, consider two neighboring vertices $v_1$ and $v_2$ and glue the corresponding patch $X_{v_1}$ on top of $X_{v_2}$ according to how $v_1$ and $v_2$ are located. Using Pachner moves we can invert the order in which $X_{v_1}$ and $X_{v_2}$ have been glued:
\begin{equation}
\label{eq:projector_commute}
\begin{gathered}
\begin{tikzpicture}
\coords{3/0,0.4|4/0.8,0.1|5/1.6,0.4|6/0.8,0.7|7/1.6,1|8/2.4,0.7};
\draw (3)edge[arr=+](4) (4)edge[arr=+](5) (3)edge[arr=+](6) (6)edge[arr=+](5) (4)edge[arr=+](6) (6)edge[arr=+](7) (5)edge[arr=-](8) (5)edge[arr=+](7) (8)edge[arr=+](7);
\shapes{\vertex}{3,4,5,6,7,8};
\end{tikzpicture}\quad\longleftrightarrow\quad
\begin{tikzpicture}
\coords{3/0,0.4|4/0.8,0.1|5/1.6,0.4|6/0.8,0.7|7/1.6,-0.2|8/2.4,0.1};
\draw (3)edge[arr=+](4) (4)edge[arr=+](5) (3)edge[arr=+](6) (6)edge[arr=+](5) (4)edge[arr=+](6) (4)edge[arr=+](7) (5)edge[arr=-](8) (5)edge[arr=-](7) (8)edge[arr=+](7);
\shapes{\vertex}{3,4,5,6,7,8};
\end{tikzpicture}\\
\Rightarrow
P_{v_1}P_{v_2}=P_{v_2}P_{v_1}
\end{gathered}
\end{equation}

So we have seen that all the linear maps $P_v$ are symmetric projectors, and they all mutually commute. If we take one copy of $X_v$ for each vertex $v$ of the physical space $S$, and glue them according to how the vertices are located in space, we end up with a triangulation of $S\times [0,1]$. The evaluation of the tensor network on this triangulation is also a symmetric projector. By construction, it is the product of the local $P_v$ projectors for all vertices of $S$ (note that the order doesn't matter as they commute).

Physically, the evaluation of the tensor-network on $S\times [0,1]$ is known as \emph{ground state projector} of a quantum system. Above we found a decomposition of this ground state projector into a product of local commuting projectors. We can thus define a local commuting-projector Hamiltonian $H$ such that the ground states of $H$ are the states in the support of the ground state projector:
\begin{equation}
H=\sum_{v \in \text{ vertices of } S} (1-P_v)
\end{equation}

Note that the construction above depends on the specific structure of $n$tSX TLs and is not generalizable to arbitrary TL types on $n$tS. In particular, Eq.~(\ref{eq:local_projector}) does not hold for more general types. For example consider the type 2tSU, for $n=2$: The tensors associated to the vertices in Eq.~(\ref{eq:local_projector}) depend on how many faces are adjacent to the vertex. But if we replace the patch $X$ on the left by the patch $X'$ on the right, the equivalent vertices on the right and left will have different numbers of adjacent vertices in the overall lattice. So the tensors on the left and right hand side are not comparable at all.

\subsection{Ground state tensor networks}
\label{sec:tensor_network_gs}
Consider a $n$tbSX TL $B$ with a $n$tSX TL $A$ as sub TL, in other words, a topological state-sum model $A$ with topological boundary $B$. For each physical space $S$ of $A$, we can construct the following patch of $B$: A triangulation $G$ of $S\times [0,1]$ where $S\times 1$ is the boundary due to cutting out the patch, with a Hilbert space associated to it, whereas $S\times 0$ is a physical boundary. We can take the triangulation to be of constant thickness in the $[0,1]$ direction, e.g. for the following space of a 2tSX TL:
\begin{equation}
\begin{gathered}
a)\quad
\begin{tikzpicture}
\coords{0/0,0|1/0.8,0.1|2/1.6,0};
\draw[ctlc1] (0)edge[arr=+](1) (1)edge[arr=+](2) (0)edge[enddots]++(-0.3,-0.1) (2)edge[enddots]++(0.3,-0.1);
\shapes{\vertex[ctlc1]}{0,1,2};
\end{tikzpicture}\quad\rightarrow\quad
\begin{tikzpicture}
\coords{0/0,0|1/0.8,0.1|2/1.6,0|3/0,0.5|4/0.8,0.6|5/1.6,0.5};
\draw (0)edge[arr=+](1) (1)edge[arr=+](2) (0)edge[enddots]++(-0.3,-0.1) (2)edge[enddots]++(0.3,-0.1);
\draw[ctlc1] (3)edge[arr=+](4) (4)edge[arr=+](5) (0)edge[arr=+](3) (1)edge[arr=+](4) (2)edge[arr=+](5) (0)edge[arr=+](4) (1)edge[arr=+](5) (3)edge[enddots]++(-0.3,-0.1) (5)edge[enddots]++(0.3,-0.1);
\shapes{\vertex}{0,1,2};
\shapes{\vertex[ctlc1]}{3,4,5};
\end{tikzpicture}\\
b)\quad
\begin{tikzpicture}
\coords{0/0,0|1/0.8,0.1|2/1.6,0};
\draw[ctlc1] (0)edge[arr=+](1) (1)edge[arr=+](2) (0)edge[enddots]++(-0.3,-0.1) (2)edge[enddots]++(0.3,-0.1);
\shapes{\vertex[ctlc1]}{0,1,2};
\end{tikzpicture}\quad\rightarrow\quad
\begin{tikzpicture}
\coords{0/0,0|1/0.8,0.1|2/1.6,0};
\draw (0)edge[arr=+](1) (1)edge[arr=+](2) (0)edge[enddots]++(-0.3,-0.1) (2)edge[enddots]++(0.3,-0.1);
\shapes{\vertex}{0,1,2};
\end{tikzpicture}
\end{gathered}
\end{equation}
a) shows a procedure to obtain a triangulation of $S\times [0,1]$ of constant ``thickness $1$''. Technically, it suffices to only take the boundary edges (yielding a patch of ``thickness $0$''), as shown in b). This is possible as the boundary can be deformed by tensor-network moves involving only the tensors associated to boundary edges on one side.

For general $n$ we can take $G$ consisting only of one boundary $(n-1)$-simplex for each $(n-1)$-simplex of $S$, as shown in b) above for $n=2$. The tensor network associated to such a patch (for any constant thickness) is a tensor-network of a geometry known as \emph{PEPS} (\emph{MPS} in one spacial dimension). E.g. for $n=2$ (with thickness $0$):
\begin{equation}
\begin{tikzpicture}
\coords{0/0,0|1/0.8,0.1|2/1.6,0};
\draw[ctlc1] (0)edge[arr=+](1) (1)edge[arr=+](2) (0)edge[enddots]++(-0.3,-0.1) (2)edge[enddots]++(0.3,-0.1);
\shapes{\vertex[ctlc1]}{0,1,2};
\end{tikzpicture}\quad\rightarrow\quad
\begin{tikzpicture}
\coords{0/0,0|1/0.8,0};
\draw (0)--(1) (0)edge[enddots]++(-0.5,0) (1)edge[enddots]++(0.5,0) (0)--++(0,0.5) (1)--++(0,0.5);
\shapes{\squarebot{180}}{0,1};
\end{tikzpicture}
\end{equation}

On the other hand, the ground state projector $P$ is the evaluation of a tensor network on a triangulation of $S\times [0,1]$ where both $S\times 0$ and $S\times 1$ are boundaries arising from cutting out the patch. If we glue the $S\times 1$ boundary of $G$ with the $S\times 0$ boundary component of the ground state projector we get a patch that is equivalent to $G$ under Pachner moves. So at the level of tensors we have:
\begin{equation}
\label{eq:ground_state}
PT(G)=T(G)
\end{equation}
where $T(G)$ is the evaluation of the tensor-network patch on $G$. Physically, Eq.~(\ref{eq:ground_state}) means that $T(G)$ is a (non-normalized) ground state of the model. So we have found that for every state-sum model with topological boundary there is a tensor-network representation of one of the ground states. More directly, the topological boundary (i.e. the tensor-network patch containing a thin stripe around the physical boundary) is the same object as the tensor-network representation. Each different topological boundary yields a different PEPS representation of a different ground state.

Thus TL types like $n$tSX TLs that automatically have a topological boundary also have tensor-network representations of a ground state. Conversely, if we take a TL type that doesn't necessarily need to have a topological boundary, like $n$tSU TLs, also this construction for a tensor-network representation fails.

\subsection{Chiral phases}
\label{sec:chiral}
Physically, $n$tS TLs are fixed point models for topological order on $n$-manifolds. In Sec.~(\ref{sec:boundary}) we saw that $n$tSX TLs can only capture phases with topological boundaries.

The topological deformability (i.e. Pachner move invariance) implies that topological models are always gapped. Thus $n$tSX TLs are fixed point models for phases which possess gapped boundaries, also known as phases ``with gappable edge'', often referred to as ``non-chiral phases''. They cannot model ``chiral'' topological phases like the quantum Hall phases, whose boundary must be gapless.

For more general $n$tS TL types (especially universal ones) however, there is no obvious construction for a standard topological (gapped) boundary. Thus it seems plausible that such universal TL types include fixed point models for phases without gappable edge. Of course the absence of a construction for a standard topological boundary for a certain TL type does not imply that there actually exist TLs for which there is no topological boundary, but it seems reasonable to assume that those exist.

There are two other indications that general TL types may be able to describe chiral phases:

1) Chiral phases are known to have no commuting-projector Hamiltonian \cite{Kapustin2018}. $n$tSX TLs do have commuting-projector Hamiltonians, see Sec.~(\ref{sec:hamiltonian}), thus they cannot describe chiral phases. However, the construction for commuting-projector Hamiltonians fails for general TL types like $n$tSU, so the latter are still possible candidates for describing chiral phases.

2) Non-chiral phases can be analyzed via tensor-network (i.e. PEPS) representations of their ground states \cite{Bultinck2015}. For chiral phases there are no known analytic tensor-network representations of the ground states comparable to the non-chiral case. It is an open question whether they have sensible tensor-network representations at all \cite{Dubail2015}. As we have seen in Sec.~(\ref{sec:tensor_network_gs}), TL types like $n$tSX have ground states with an exact analytic tensor-network representation, which are equivalent to those in \cite{Bultinck2015}. However, the construction for tensor-network representations of ground states fails for general TL types like $n$tSU, so the latter are still consistent with chiral phases.

\section{Outlook}
\label{sec:outlook}
In this paper we suggested a generalization of state-sum constructions and introduced the concept of universality for those generalizations. We do believe that there are instances of the generalized state-sum construction that provide fixed point models for phases, which are not captured by conventional state-sum constructions. So far we haven't found a concrete example for such a novel fixed point model though. In this section we describe a few ideas which might lead to concrete examples of new fixed point models.

Roughly we have the following approaches:
\begin{enumerate}
\item Try to find new instances of the generalized state-sum constructions numerically.
\item Find a universal TL type with a simpler set of axioms.
\item Look at generalized state-sum constructions for other lattice classes or other tensor types.
\item Search the literature for algebraic structures that satisfy axioms which are similar to the axioms of the generalized state-sum constructions, and then construct TLs from examples of this algebraic structure.
\end{enumerate}

In the following sections we discuss ideas along these lines.

\subsection{Numerical search for fixed point models}
The axioms for a TL with real tensors are a set of equations between different tensor networks formed by those tensors. These are multi-variable polynomial equations. Finding roots of these polynomial equations can be done by e.g. using a Gauss-Newton method. Surely the computational cost of finding roots grows exponentially with the number of open indices in the axioms. This is not a fundamental problem, as the number of open indices is fixed and doesn't scale for a specific state-sum construction. However, it makes finding solutions for the universal TL types described in this paper practically intractable for dimensions $n\geq 3$, as the corresponding tensor-network equations are quite complicated and have a large number of open indices.

A possibility that could make dealing with large contractions of tensors computationally more feasible is to restrict to a subset of tensors that can be efficiently contracted (in the number of indices), such as stabilizer tensors or Gaussian tensors \cite{tensor_type}. Of course, such a restriction would also decrease the amount of different phases that we can possibly represent.

\subsection{Simpler set of axioms}
The universal state sum construction we proposed is already rather simple compared to ad hoc constructions that one could think of. It still suffers from a large number of tensors and axioms, involving a large number of open and contracted indices. There might be alternative equivalent sets of tensors and axioms that are less systematic but involve a smaller number of indices. It seems to be the case that for such a construction we need a much more refined construction where the proof of universality is more tricky.

\subsection{Projective tensors}
\label{sec:projective}
Physically, TL tensor networks can be probed by inserting tensors with open indices. Then the evaluation of such a tensor network yields a tensor which is a probability distribution describing the frequencies of measurement outcomes. By construction, these probability distributions are normalized. So if two TLs only differ by local normalization factors they are physically indistinguishable.

Moreover, if the TL represents an imaginary time evolution of a quantum model in euclidean space-time (which it does when considering ground state quantum phases), then the actual physical tensor-network that is probed by inserting measurement tensors consists of two (complex-conjugated) copies of the TL stacked together. This doubled tensor-network doesn't change at all under phase prefactors added locally to the tensors. So again the physics doesn't care for local prefactors.

A way to formalize this physical insight within our framework is to use a different tensor type called \emph{projective tensors}. These are (real or complex) tensors modulo multiplying with a (real or complex) number, i.e. either a ray in a vector space or the zero vector. Using projective tensors instead of real or complex tensors has the consequence that the tensor-network equations arising from Pachner moves only hold up to a scalar prefactor, allowing for more general solutions.

It seems that in order to get TLs for the most general physical phases it is necessary to use projective tensors. Indeed, most topological phases in $2+1$ dimensions have a local continuum description in terms of Chern-Simons theory \cite{Witten1989}. For ``chiral'' models, the latter has a so-called anomaly meaning that the partition function on a manifold does not only depend on its topology, but on an additional structure such as a framing. Changes of the framing are accompanied by a phase connected to the topological central charge of the quantum group that is input to the construction. This suggests that a discrete version of chiral Chern-Simons theory should obey retriangulization invariance only projectively.

A large part of the literature on state-sum constructions is motivated by finding manifold invariants. Projective tensors cannot yield non-trivial manifold invariants, as there are only two projective scalars, represented by $0$ and $1$. This is why projective constructions haven't been considered much in the literature (apart from e.g. axiomatic ($1$-extended) TQFTs, where using projective tensors is known as ``anomaly''). However, manifold invariants themselves are unphysical, as the partition function on a closed manifold is clearly not a quantity that is directly measurable.

One can map real or complex tensors to projective tensors taking the corresponding equivalence class under multiplication by local prefactors. This map is consistent with tensor product and contraction. We call such maps \emph{tensor mappings}.

\subsection{Universality for other lattice classes}
In this paper we introduced the concept of universality for topological state sums. However, universality can be formalized as a much more general concept. For every pair of lattice type and tensor type, a TL type is universal if it can emulate any other TL type with the same lattice type and tensor type. Thus, it is possible to give universal TL types corresponding to, e.g., topological boundaries, domain walls, or anyons, or also phases with fermions or protected by symmetries.

The construction of the universal type $n$tSU can be generalized to arbitrary \emph{higher order cell complexes} \cite{tensor_lattice}: Put one tensor on each $x$-cell, such that every pair of adjacent $x$-cell and $(x+1)$-cell shares a contracted index. The tensor depends on the \emph{region} the $x$-cell is part of, as well as on its \emph{upper} and \emph{lower link}. With such a construction it is in principle possible to obtain state-sum models for, e.g., anyons, or domain walls beyond the Drinfel'd center, or bi-modules of fusion categories.

\subsection{Fixed point models for new unphysical phases}
In this section we consider real or complex TLs consisting of only scalars, i.e. tensors without indices, or only trivial bases (namely the one-element set). Such TLs are physically irrelevant according to Sec.~(\ref{sec:projective}), as they become trivial after applying the tensor mapping to projective tensors. If we disregard physical relevance, there are indeed TL phases that are representable by universal TL types, but not by conventional state-sum constructions. Even though these examples for phases are physically trivial, they do give hope that there might also be new physical phases described by universal TL types.

If we interpret the all-scalar TLs as the imaginary time evolution of a quantum model, their physical triviality also can be seen in the following way: The physical spaces correspond to co-dimension $1$ cuts through the space-time tensor network. The degrees of freedom constituting the many-body Hilbert space are given by all the contracted indices divided by the cut. All-scalar TLs don't have any contracted indices, so they correspond to physical models without any degrees of freedom (i.e. whose Hilbert space is $1$-dimensional). Such models are obviously trivial.

The Euler characteristic $\chi(X)$ is a topological invariant of a $n$-dimensional simplicial complex $X$ which can be formulated as an all-scalar state sum, in the following sense: For every (real or complex) $\alpha$, there is a (real or complex) $n$tSU TL whose evaluation on $X$ yields $\alpha^{\chi(X)}$. This $n$tSU TL consists of only tensors which are scalars. It associates the number $\alpha^{(-1)^x}$ to each $x$-simplex, i.e. $\alpha$ to each vertex, $\alpha^{-1}$ to each edge, and so on. Evaluated on a simplicial complex $X$ with $\#x$ $x$-simplices, and so on, we get:
\begin{equation}
\prod_{0\leq x \leq n} \alpha^{(-1)^{x}\#x}=\alpha^{\sum_{0\leq x \leq n}(-1)^{x}\#x}=\alpha^{\chi(X)}
\end{equation}
Note that this is only non-trivial for even $n$, since otherwise $\chi(X)$ is always $0$. For each $\alpha$, this TL is in a different phase, as there are manifolds on which the topological invariants obtained by evaluating the tensor networks are different.

This kind of construction cannot be directly represented by the TL type $n$tSX, as the latter doesn't have tensors associated to the vertices and faces. In fact, at least in $2$ dimensions, there is only a discrete set of values of $\alpha$ such that there is a $n$tSX TL in the same phase.

Another set of ($\mathbb{Z}_2$) topological invariants is given by the Stiefel-Whitney numbers. For each decomposition of $n=n_0+\ldots+n_i$ the corresponding Stiefel-Whitney number is given by the integral over the cup product of the Stiefel-Whitney obstruction classes $\omega_{n_0},\ldots,\omega_{n_i}$. It can be computed combinatorially for a simplicial complex $X$ in the following way: 1) The Poincar\'e dual of $\mathbb{Z}_2$ $x$-cocycles on a $n$-manifold $X$ can be represented by $(n-x)$-cycles, which are collections of $(n-x)$-simplices of $X$ with empty boundary. There are combinatorial formulas that compute a $(n-x)$-cycle in $X$ corresponding to a representative of the $x$th Stiefel-Whitney obstruction class, given e.g. in \cite{Goldstein1976}. These formulas are local in the sense that whether a $(n-x)$-simplex is contained in the cycle or not only depends on a constant-size neighborhood of the simplex. 2) The cup product of a $x$-cocycle and a $y$-cocycle is a $x+y$-cocycle. There is a combinatorial formula that computes the cup product of the corresponding $(n-x)$-cycles and $(n-y)$-cycles yielding a $(n-x-y)$-cycle, given e.g. in \cite{Steenrod1947}. (Note though that in \cite{Goldstein1976} the formulas yield cycles in the simplicial complex, whereas the formulas in \cite{Steenrod1947} are for cycles in its Poincar\'e dual complex.) Also this formula is local in the same sense. 3) Integration over a $n$-cocycle corresponds to ($\mod 2$) counting of $0$-simplices of the corresponding $0$-cycle (or $n$-simplices in the dual formulation).

Combining the local combinatorial formulas for 1) and 2) above we get a local combinatorial formula that computes a collection of vertices of $X$ representing the cup product of $\omega_{n_0},\ldots,\omega_{n_i}$. From this we can construct a TL by associating the number $-1$ to all vertices in the collection. Evaluating this TL on a simplicial complex $X$ yields $(-1)^{S(X)}$ where $S(X)$ is the Stiefel-Whitney number associated to $X$ (or the underlying manifold). Again this TL phase cannot be represented by simple TL types like $n$tSX, but the tensor-network can be reshaped to a TL of type $n$tSU.

\subsection{The Crane-Yetter-Walker-Wang models}
As we motivated in Sec.~(\ref{sec:chiral}), a plausible candidate that could be described by our generalized state-sum construction are chiral phases in $2+1$ dimensions. As discussed in Sec.~(\ref{sec:projective}) one should use complex projective tensors to capture these phases due to their framing anomaly. In this section we will sketch a strategy that might lead to an intrinsically $2+1$-dimensional state-sum for (possibly) chiral phases, taking a modular fusion category as input.

We will start with a $3+1$-dimensional topological state-sum construction known as the Crane-Yetter model \cite{Crane1993} which was later put into a Hamiltonian formulation known as the Walker-Wang model \cite{Walker2011} (thus referred to as \emph{CYWW model}). This model takes a unitary braided fusion category as input and yields the following:
\begin{itemize}
\item A $3+1$-dimensional topological state-sum construction.
\item A state-sum construction for a $2+1$-dimensional topological boundary of the latter.
\item For each object in the fusion category, a $0+1$-dimensional ribbon defect within the latter boundary.
\end{itemize}
Consider a fixed $3$-manifold, with different ribbon graphs, and different $4$-manifolds bounding the $3$-manifold. The correlation functions of the CYWW model on the $4$-manifold with ribbon boundary are equal to those of the Reshetikhin-Turaev TQFT on the ribbon boundary, for the same modular fusion category, apart from a phase factor which only depends on the bulk $4$-manifold (not on the ribbon graph).

For our purposes we restrict to input unitary braided fusion categories that are modular. In this case the $3+1$-dimensional model is known to be invertible and to only have trivial ground state degeneracy and no quasi-particles \cite{Walker2011}. The corresponding state-sum is not immediately physically trivial in the sense of the Sec.~(\ref{sec:projective}), as it has non-scalar tensors with non-trivial contracted indices. However, we will show in the next paragraph that there is a all-scalar state-sum that produces the same topological invariants as the CYWW state-sum. This makes it highly plausible that the CYWW state-sum and this all-scalar state-sum are in the same phase.

The partition function of the CYWW model on a $4$-manifold $X$ equals $e^{ic\pi \sigma(X)/4}$ where $c$ is the topological central charge of the input modular fusion category and $\sigma(X)$ is the signature of $X$. According to the Hirzebruch signature theorem, $\sigma(X)=P(X)/3$ where $P$ is the integral of the Pontryagin class of $X$. Just as for euler or Stiefel-Whitney classes there exist local combinatorial formulas for Pontryagin classes. This yields a complex all-scalar TL reproducing the CYWW invariants. More precisely, this all-scalar TL is defined on a combinatorial version of oriented manifolds, and whether we take a scalar or its complex conjugate depends on this combinatorial orientation of the simplicial complex.

Now consider the CYWW model with boundary and apply the disentangling circuit that transforms the CYWW bulk to the all-scalar TL from the paragraph above. After going to projective tensors, the TL inside the $4$-dimensional bulk becomes trivial and we're left with a genuinely $3$-dimensional TL (with ribbon defects) on the boundary. So this procedure yields a state-sum fixed point model for any possible anyon theory (given by the modular fusion category). This comes at the expense of having to use projective tensors though.

\subsection*{Acknowledgments}
We thank C. Wille, A. Nietner and J. Eisert for lots of discussions and proof reading the document. We thank the DFG (CRC 183, B01) for support.

\bibliography{universal_state_sums_refs}{}
\bibliographystyle{utphys}

\end{document}